\documentclass[aps,english,prl,floatfix,superscriptaddress,showpacs,amsmath,amssymb,reprint]{revtex4-2}
\usepackage[T1]{fontenc}
\usepackage[latin9]{inputenc}
\usepackage{babel,booktabs}
\usepackage[unicode=true]{hyperref}
\usepackage{graphicx}

\makeatletter
\usepackage{bbold}
\usepackage{dsfont}
\PassOptionsToPackage{caption=false}{subfig} 
\usepackage{hyperref}
\usepackage{feynmf}
\hypersetup{
breaklinks=true,
colorlinks=true,
citecolor=blue,
linkcolor=blue,
filecolor=blue,
urlcolor=blue
}

\IfFileExists{lmodern.sty}{\usepackage{lmodern}}{}

\newcommand{\dg}{^{\dagger }}

\newcommand{\bk}{{\bf{k}}}

\newcommand{\rarrow}{\rightarrow}

\makeatother

\begin{document}
\title{Mobile Majorana zero-modes in two-channel Kondo insulators}
\author{Milan Kornja\v ca }
\affiliation{Department of Physics and Astronomy, Iowa State University, Ames, Iowa 50011, USA}

\author{V. L. Quito} 
\affiliation{Department of Physics and Astronomy, Iowa State University, Ames, Iowa 50011, USA}

\author{R. Flint}
\email{flint@iastate.edu}
\affiliation{Department of Physics and Astronomy, Iowa State University, Ames,
Iowa 50011, USA}

\date{\today}
\maketitle
\textbf
{Non-abelian anyons are highly desired for topological quantum computation purposes, with Majorana fermions providing a promising route, particularly zero modes with non-trivial mutual statistics. Yet realizing Majorana zero modes in matter is a challenge, with various proposals in chiral superconductors, nanowires, and spin liquids, but no clear experimental examples. Heavy fermion materials have long been known to host Majorana fermions at two-channel Kondo impurity sites, however, these impurities cannot be moved adiabatically and generically occur in metals, where the absence of a gap removes the topological protection. Here, we consider an ordered lattice of these two-channel Kondo impurities, which at quarter-filling form a Kondo insulator.  We show that topological defects in this state will host Majorana zero modes, or possibly more complicated parafermions.  These states are protected by the insulating gap and may be adiabatically braided, providing the novel possibility of realizing topological quantum computation in heavy fermion materials.}

Majorana fermions ($\gamma$) are real fermions that are their own antiparticles.  These may be free particles,
but are most interesting when bound to defects as a zero energy state in two-dimensions~\cite{dassarmareview}.  These defects occur in pairs, which may be spatially separated,
with the two Majoranas encoding a single complex fermion, $c\dg = \gamma_1 + i \gamma_2$.  The resulting ground state degeneracy is associated with the complex fermion parity, $c\dg c = 0,1$, and can encode qubits.  If two defects can be braided adiabatically, the system evolves smoothly within the degenerate ground state manifold, encoding non-Abelian mutual statistics, even as the self-statistics remain fermionic. These defects are called Majorana zero modes (MZMs) to distinguish them from free Majoranas, and are Ising anyons.  The encoded quantum information can be controlled by manipulating the defects and is topologically protected by the system gap. MZMs require only that the fermions are real, the defect state is localized, and the fermion parity is conserved.  Many proposals use superconductivity to ensure parity conservation \cite{Kitaev_2001,Ivanov_PRL_2000,Alicea_NatPhys_2011,DasSarma_PRL_2010,Mourik_Science_2012}, but the emergent fermions in spin liquids, and in some heavy fermions, as we shall see, are also viable candidates \cite{Kitaev_AnnPhys_2006,Banerjee_Nat_Mat_2016,Kitaev_AnnPhys_2002}.

\begin{figure}[!htb]
\includegraphics[width=0.95\columnwidth]{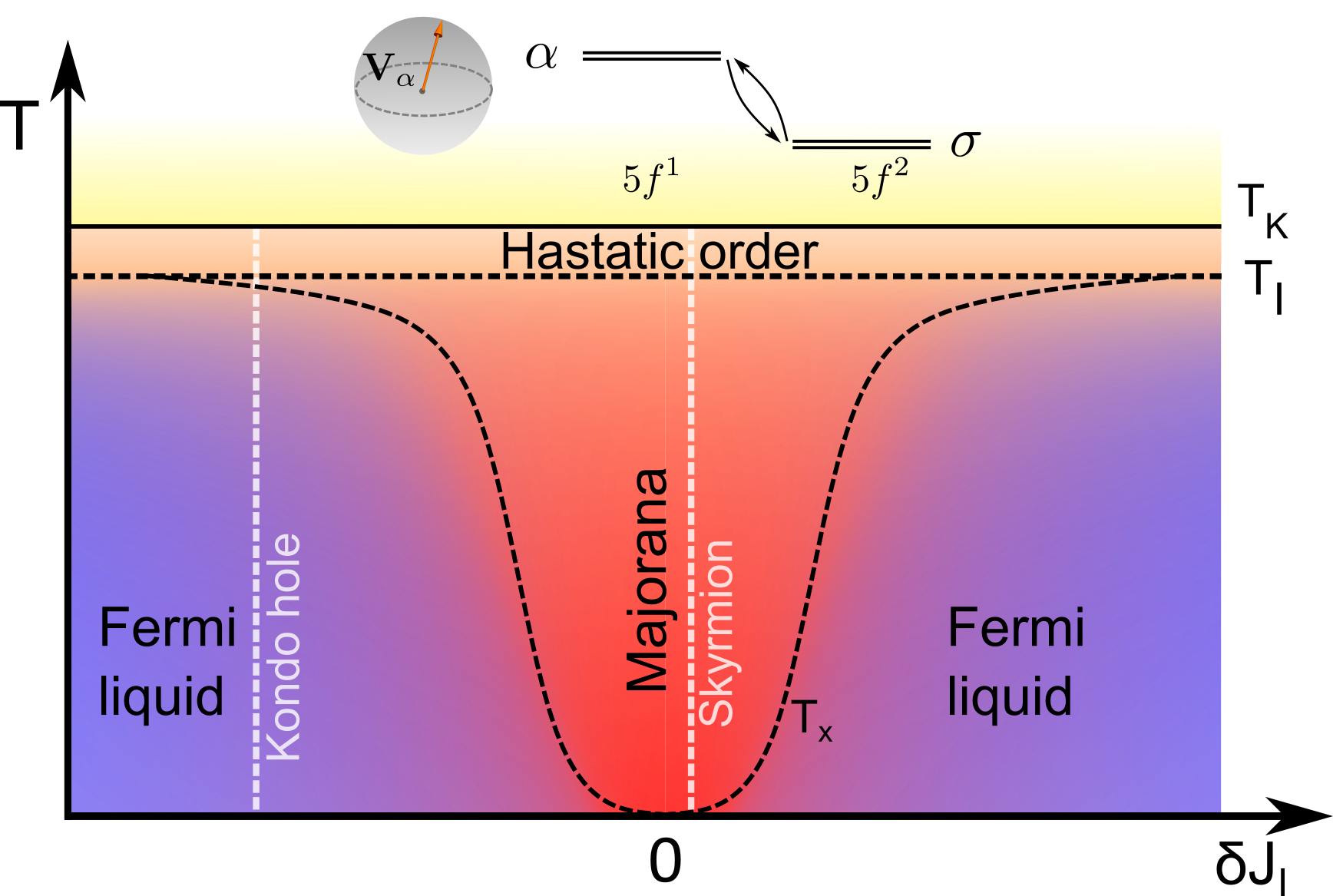}
\caption{\textbf{Phase diagram of defects in two-channel Kondo lattices in channel asymmetry, $\delta J_I$ and temperature, $T$.}  Top: Hastatic order arises from valence fluctuations between doublet ground and excited states, leading to two-channel Kondo physics. The finite hybridization ($\mathbf{V}_\alpha$) breaks SU(2) channel symmetry. (Bottom) We explore two types of defects, Kondo holes (point defects) and skyrmions (topological defects). The Kondo insulating phase transition is shown with a full line; dashed black lines represent crossovers.  The defect states lead to emergent Kondo impurity physics. In the case of the Kondo hole, $\delta J_I$ is large and a single-channel local Fermi liquid is formed. The skyrmion shows almost perfect channel symmetry, leading to two-channel Kondo impurity physics, thus hosting Majorana zero modes for temperatures below the impurity Kondo scale ($T_I$) and above the channel asymmetry crossover scale ($T_x \sim \delta J_I^2$). These Majoranas have non-trivial mutual statistics and are mobile, making them good candidates for topological quantum computation. \label{fig:Fig1}}
\end{figure}
Kondo impurities are isolated local moments that interact antiferromagnetically with a metallic host,
and are screened by the conduction electron spin density at low temperatures.  
Usually, this screening involves a single channel of conduction electrons, but sometimes the crystal symmetry guarantees two or more symmetry-related channels~\cite{NozieresBlandin_JPhys_1980}.  This \emph{multi-channel} Kondo interaction \emph{overscreens} the local moment, leading to a non-Fermi liquid with fractional impurity ground state  entropy~\cite{NozieresBlandin_JPhys_1980,parcollet}.  Two-channel Kondo impurities have a zero-point entropy of $\frac{1}{2} R\ln 2$ and host a single Majorana fermion~\cite{Kivelson1992}.  These impurities can arise naturally in materials with 4f$^2$ or 5f$^2$ impurities in high symmetries, like cubic or  tetragonal, such that the ground state is a \emph{non-Kramers} doublet protected by crystal rather than time-reversal symmetry~\cite{NozieresBlandin_JPhys_1980,Cox1998}.  The Kondo effect is driven by valence fluctuations to excited $f^{1}$ or $f^3$ states, which are Kramers doublets, guaranteeing two degenerate screening channels, as shown in the inset of Fig.~\ref{fig:Fig1}.  Experimentally, two-channel Kondo scaling behavior has been observed in dilute systems~\cite{amitsuka1994},
but the zero point entropy associated with Majoranas is not found~\cite{toth}, presumably due to the absence of a gap.  More importantly, these impurities cannot be moved adiabatically.

An ordered lattice of two-channel Kondo impurities can host rich physics, from magnetism and superconductivity to non-Fermi liquids~\cite{Cox1998} and an unusual channel-symmetry breaking heavy Fermi liquid called \emph{hastatic order}~\cite{Hoshino2011,Chandra2013}. Here, we take the conduction electrons to be quarter-filled, in which case the ground state of the two-channel Kondo lattice for sufficiently strong Kondo coupling is known to be a channel antiferromagnetic (antiferrohastatic) Kondo insulator~\cite{DMRG,Zhang2018}.  

Defects in this insulating state can give rise to localized in-gap doublets that interact with the surrounding heavy bulk states via a two-channel Kondo interaction that converts them into MZMs.  We examine both Kondo holes, where a single local moment is removed, and hybridization skyrmions, which are stable, and mobile, topological defects~\cite{Wugalter2019}.  Both defects locally break the channel symmetry, but Majorana behavior is still expected for sufficiently small symmetry breaking~\cite{rozhkov1998impurity,Cox1998,LeHur2016},
as is generically the case for the skyrmion, and is sketched in Fig.~\ref{fig:Fig1}.

{\bf Two-channel Kondo insulators}

Our starting point is a two-channel Kondo lattice, where the local moments are pseudospin $\frac{1}{2}$, $\vec{S}_{f,j}$ that are screened by conduction electron pseudospins, $\vec{s}_{c,j,\alpha} = \sum_{\sigma,\sigma'} c\dg_{j \alpha \sigma} \vec{\sigma}_{\sigma, \sigma'} c_{j \alpha \sigma'}$ in two distinct channels labeled by $\alpha = 1,2$.  The $SU(2)$ pseudospin degree of freedom is labeled by $\sigma = \pm$.  We choose the conduction electrons to be quarter-filled, which leads to an antiferrohastatic ground state with a hybridization gap opening at the Fermi level \cite{Zhang2018} that we can capture within a simple mean-field theory. The Kondo lattice Hamiltonian is,
\begin{equation}
    H = \sum_{\bk \alpha \sigma}\epsilon_{\bk} c\dg_{\bk \alpha \sigma} c_{\bk \alpha \sigma} + J_K \sum_{j,\alpha \sigma \sigma'} c\dg_{j \alpha \sigma} \vec{\sigma}_{\sigma \sigma'} c_{j \alpha \sigma'} \cdot \vec{S}_{f,j}.
\end{equation}
To proceed, we take the $SU(N)$ large-$N$ limit, where this model can be solved exactly~\cite{Coleman_PhysRev_1983,Chandra2013,Zhang2018,Wugalter2020}, with the results expected to apply well to the $N = 2$ realistic limit
Here, the local moment pseudo-spins, $\vec{S}_{f,j}$ are promoted to $SU(N)$ moments and represented by pseudofermions, $\vec{S}_{f,j} = \sum_{\sigma \sigma'}f\dg_{j\sigma}\vec{\sigma}_{\sigma \sigma'}f_{j\sigma'}$, where $\vec{\sigma}$ is a generator of $SU(N)$.  This representation faithfully captures the spin physics if the pseudofermions are exactly half-filled on every site, which is enforced by a local constraint field, $\lambda_j$. The average, $\lambda$ effectively acts as a chemical potential for the pseudofermions.

The resulting quartic Hamiltonian can be decoupled to give a quadratic Hamiltonian with channel dependent mean-field hybridizations, $V_{j\alpha} = \frac{J_K}{N}\langle \sum_\sigma f\dg_{j\sigma}c_{j\alpha\sigma}\rangle$ on every site~\cite{Chandra2013,Zhang2018}. 
The mean-field Hamiltonian (exact at $N=\infty$) is then,
\begin{equation}
    H = \sum_{\bk \alpha \sigma}\epsilon_{\bk} c\dg_{\bk \alpha \sigma} c_{\bk \alpha \sigma} + \sum_{j \alpha \sigma} V_{j\alpha} c\dg_{j \alpha \sigma} f_{j \sigma } +\mathrm{H.c.}+\sum_j\frac{N |{\bf V}_j|^2}{J_K}.
\end{equation}
These hybridizations are the hastatic order parameter, and are fundamentally a \emph{spinorial} quantity that condenses below a phase transition at the Kondo temperature, $T_{K}$,
\begin{equation}
{\bf V}_{j}=|V|e^{i\varphi_j/2}\left(\begin{array}{c}
e^{i\phi_j}\cos\frac{\theta_j}{2}\\
\sin\frac{\theta_j}{2}
\end{array}\right),
\end{equation}
where the two components are $\alpha = 1,2$ and the spatial arrangement is governed by the lattice details.

This spinorial hybridization always breaks the $SU(2)$ channel symmetry, with most of the physics captured by a vectorial composite order parameter that corresponds to the magnetic moment of the on-site hybridization, ${\bf V}_j\dg\vec{\sigma}{\bf V}_j$ and has an $SO(3)$ symmetry~\cite{Hoshino2011}.  The spinorial nature can result in distinct arrangements that differ only by phase differences, $\varphi_i - \varphi_j$ 
and break different symmetries while having the same moment arrangement~\cite{Zhang2018,Kornjaca2020}.

In addition to the Kondo interaction, we also consider $f$ hopping terms.  These are an emergent phenomenon naturally arising above the ordering temperature for any finite $N$.  Traditionally, these are theoretically obtained in the large-$N$ limit by adding an antiferromagnetic RKKY coupling of the moments, 
$J_H \vec{S}_{f,i}\cdot \vec{S}_{f,j}$, and decoupling the resulting quartic terms as above.  This decoupling leads to mean-field hoppings, $\chi_{ij} = \frac{J_{H}}{N}\langle \sum_\sigma f\dg_{i\sigma} f_{j\sigma}\rangle$ that give an additional term $H_f = \sum_{\boldsymbol{k},\sigma}\epsilon_{f\bk} f\dg_{\bk \sigma}f_{\bk \sigma}$.  These hoppings are generically present, differentiate spinorial orders with differing phases, and ensure the existence of topological defect states.

\textbf{Antiferrohastatic square-octagon model}
\begin{figure}
\includegraphics[width=0.95\columnwidth]{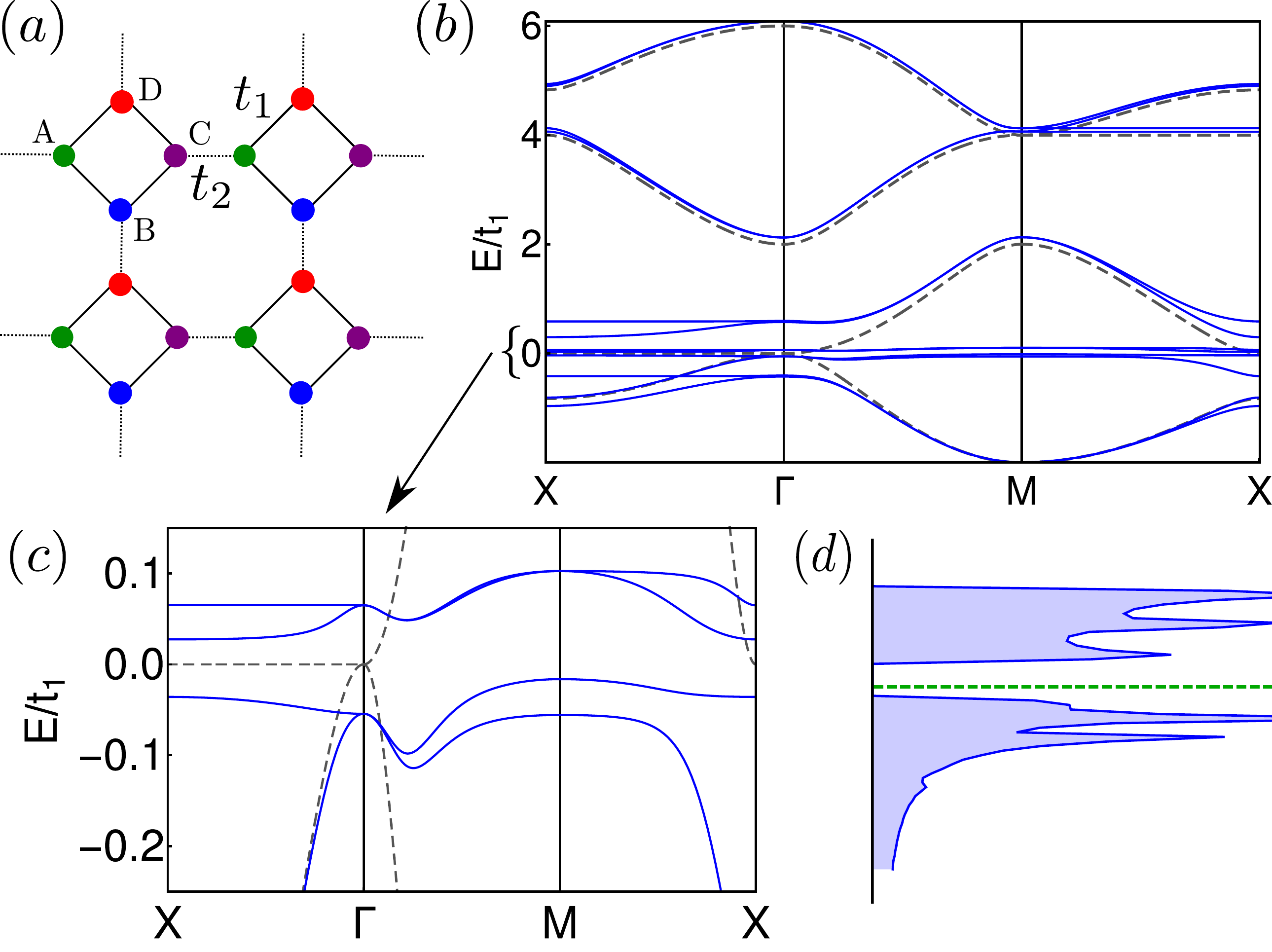}
\caption{\textbf{Square-octagon lattice and hybridized bands} (a) Square-octagon lattice, with four sites per unit cell and hoppings $t_{1}$ and $t_{2}$. (b) Dispersion along a high symmetry path in the Brillouin zone for $t_2 = 2 t_1$. The two-fold degenerate unhybridized conduction bands (dashed gray) present a quadratic band touching at the $\Gamma$ point. The hybridized and gapped antiferrohastatic bands are shown in blue (full lines). (c) Bands close to $E=0$ showing the gap. In (d), the corresponding density of states. The dashed green line is the position of the skyrmion defect state. See Methods for parameter values.
\label{fig:Fig2}}
\vspace{-0.5cm}
\end{figure}

Motivated by the stability of the Kondo insulator in the quarter filled square lattice~\cite{Zhang2018} and wanting the single in-gap defect state that arises from a point-like Fermi surface~\cite{Senthil2003}, we consider a toy model on a square-octagon lattice (also known as 1/5-depleted or CAVO lattice)~\cite{KatohImada_JJP_1995}. The four sites per unit cell, shown in Fig.~\ref{fig:Fig2}(a) allow a quadratic band touching (QBT) at the $\Gamma$ point exactly at quarter filling for a wide range of parameter choices~\cite{Fiete_PRB_2010}.

We are interested in the staggered, or ``antiferrohastatic'' (AFH) hybridization arrangement that preserves a time-reversal-like anti-unitary symmetry. This is the ground state at the quarter filling, where it generically opens up a full hybridization gap~\cite{Zhang2018}. Due to the spinorial nature of hastatic order, a four sublattice ansatz (4SL) is needed in order to preserve a time-reversal-like symmetry. We choose the simplest such ansatz on the square-octagon lattice where the composite moments alternate within the unit cell:
\begin{align} \label{eq:4sl}
{\bf V}_{j,B}&=\theta{\bf V}_{j,A},\qquad {\bf V}_{j,C}=\theta^{2}{\bf V}_{j,A}=-{\bf V}_{j,A},\nonumber \\
{\bf V}_{j,D}&=\theta^{3}{\bf V}_{j,A}=-{\bf V}_{j,B},
\end{align}
and this pattern is repeated in other cells such that the unit cell is not further enlarged. This 4SL state returns to itself after a combination of time-reversal and a local rotation ($A \rarrow B \rarrow C \rarrow D \rarrow A$). Note, the actual ground state expected from a strong coupling analysis ($J_K\gg t$) is the ground state of an effective Heisenberg model for the composite moments~\cite{Cox1998,Zhang2018}, for which this arrangement would be staggered between unit cells~\cite{Manuel_PRB_1998}. However, this arrangement requires doubling the number of bands and the key aspects of the physics are already captured by the simpler state. The results should be generic to any state that fully gaps out the Fermi surface with an antiferrohastatic ground state preserving a time-reversal-like symmetry.

In this mean-field model, a gap opens up at the $\Gamma$ point for sufficiently strong Kondo coupling.  The chemical potential, $\mu$ and constraint, $\lambda$ adjust self-consistently to keep conduction electrons at quarter-filling and the $f$-electrons at half-filling, such that the gap opens around the Fermi level, leading to a Kondo insulator with heavy bands below and above the gap, as shown in Fig.~\ref{fig:Fig2}(b)-(d).  We fix $|V|$ and $\chi$, which corresponds to some set of Kondo/Heisenberg couplings.  The resulting hybridized bandstructure is complicated due to the large number of bands, and while the time-reversal-like symmetry guarantees Kramers degenerate pairs, they are located at different points in momentum space. As we will see, this symmetry guarantees two degenerate heavy electron channels and is weakly broken by the defects.

\textbf{Defects: Kondo holes and skyrmions}

\begin{figure}
\includegraphics[width=0.95\columnwidth]{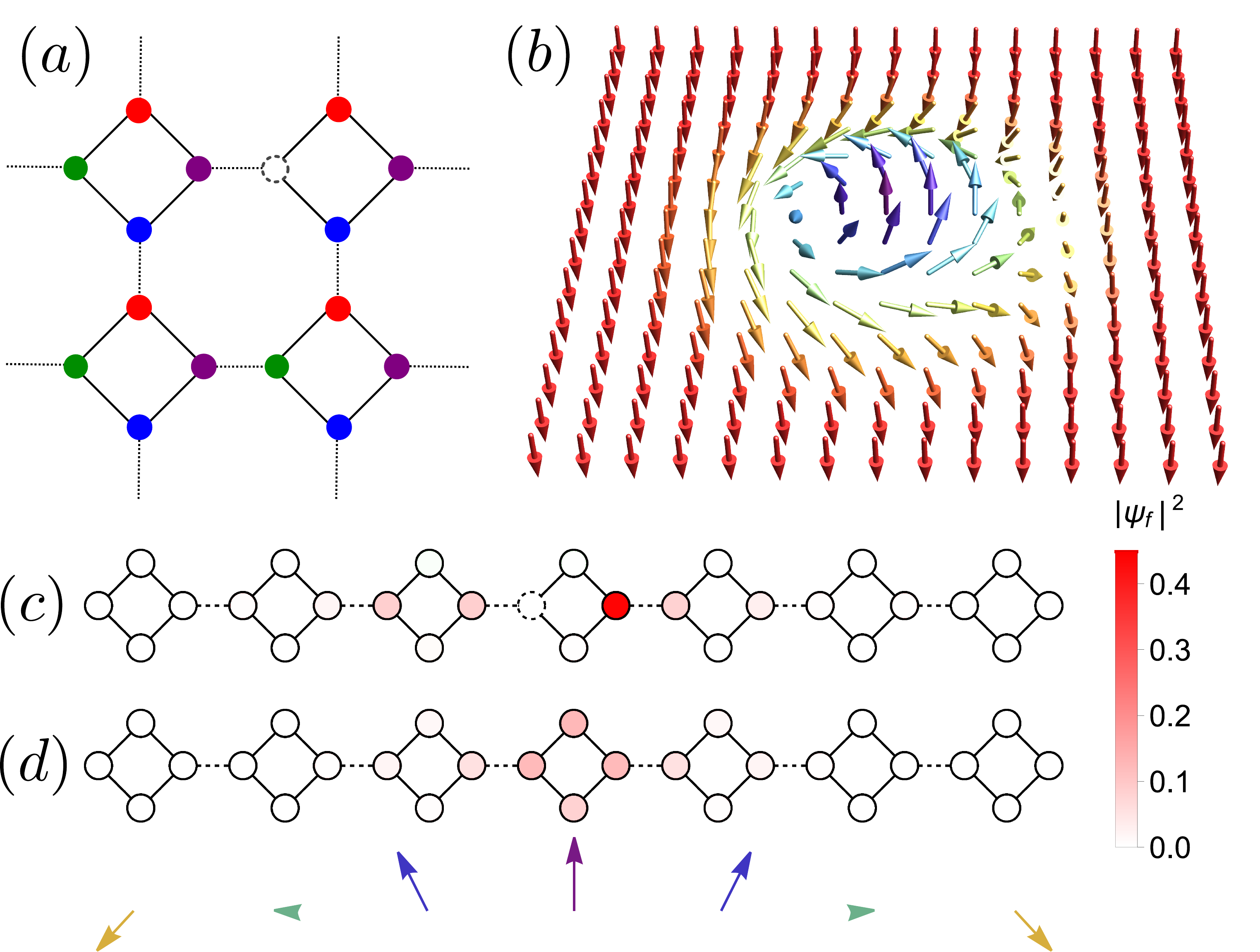}
\caption{\textbf{Defects in two-channel Kondo lattices} (a) Kondo hole defect at an $A$ site. (b) Topological hybridization skyrmion defect. Arrows show the direction of hastatic spinor on the $A$ site of each square-octagon unit cell. Spinors on other sublattices follow the 4SL AFH pattern. (c)-(d) Wave function amplitude of the Kondo hole and skyrmion defect states on a strip of seven unit cells along the horizontal ($[10]$) direction, around the center cell. The magnitude of the $f$-electron wave function projection on each site is shown by the color intensity. In (d), the skyrmion spinor projections on the $z$-axis are shown at the bottom. See Methods for more details and parameter values. \label{fig:Fig3}}
\vspace{-0.5cm}
\end{figure}

We now examine defects in this Kondo insulating state, treating two distinct types that showcase different behavior. The first is a Kondo hole: the absence of a local moment on one lattice site, shown in Fig.~\ref{fig:Fig3}(a). In the single-channel case, the Kondo hole hosts an orthogonal localized bound state~\cite{Schlottmann1991,Figgins2011}, which exhibits an impurity single-channel Kondo effect~\cite{Tesanovic1986,Schlottmann1991}. A critical behavior analogous to the two-channel Kondo impurity might be expected, but is not guaranteed. The second type is a hybridization (or channel) skyrmion, a topological defect in the hastatic order parameter in which the composite moments form a two-dimensional antiferromagnetic skyrmion~\cite{Mermin1979,Wugalter2020}.  
As the moments vary smoothly, the skyrmions are weakly pinned to the lattice and may be easily moved, unlike Kondo holes. As the composite moments are tiny magnetic moments \cite{Chandra2013,Zhang2018}, these hybridization skyrmions will couple to conventional magnetic skyrmions in a thin film geometry, allowing the generation and manipulation of these defects.

We introduce the Kondo hole state by raising the local $f$-electron chemical potential, $\lambda_0 \rarrow \infty$ on the central $A$ site, and suppressing the hybridization on this site, $|V|_0 = 0$.  For all other sites, we use the parameters for the translationally invariant system~\cite{Schlottmann1991}.  The full real-space self-consistent solution for a Kondo hole does not significantly change the  parameter values in a heavy fermion metal~\cite{Figgins2011}, so we expect this approximation to be reasonable, particularly for a Kondo insulator. 

For simplicity, we consider skyrmions where the composite moments within a unit cell remain aligned/anti-aligned as in Eq.~\eqref{eq:4sl}, with the A moments forming the skyrmion shown in Fig.~\ref{fig:Fig3}(b).  The magnetization points ``up'' within a core region, defined by the skyrmion radius, $R_0$, and smoothly distorts to point ``down'' far from the core, with the in-plane components winding around the core, as in a vortex, with the mathematical expression given in the Methods. Here, we vary only the hybridization \emph{direction} at each site, leaving $|V|$ and $\lambda$ uniform.

We treat the defect states explicitly by diagonalizing the mean-field Hamiltonian in real space on finite lattices (see Methods for more details). 
For both defects, pseudospin degenerate bound states are found in the Kondo insulating gap at the Fermi level [see Fig.~\ref{fig:Fig2} (d)], where they are half-filled. The defect states are $f$ in character and highly localized, 
as shown in Fig.~\ref{fig:Fig3}(c)-(d). The exponential decay is governed by a localization length related to the inverse of the gap. Close to the defect center, the Kondo hole strongly breaks lattice symmetry, with the defect state mostly localized on the C site of the same unit cell and orthogonal to the $f$-hole itself. By contrast, the skyrmion breaks lattice symmetry only weakly.  This distinction in local symmetry breaking ultimately leads to very different channel asymmetries for their two-channel impurity Kondo coupling.

The numerical diagonalization gives a mean-field Hamiltonian with both heavy bulk and impurity states,
\begin{align}
    H_{MF}=\sum_{\kappa,\eta,\sigma}E_{\kappa\eta}\beta_{\kappa\eta\sigma}^\dagger\beta_{\kappa\eta\sigma}+E_I \beta_{I\sigma}^\dagger\beta_{I\sigma},
\end{align}
where $I\sigma$ denotes the localized defect zero-energy doublet and its corresponding energy, $E_{I} = 0$, while $\kappa\eta\sigma$ labels the heavy conduction electron states. The emergent fermionic operators, $\beta$, are linear combinations of the $c$ and $f$ fermions.  $\sigma$ continues to denote the pseudospin, while $\kappa$ replaces the crystal momentum in the translationally invariant case, and $\eta$ is a channel index denoting two approximately degenerate bulk states for each $\kappa\sigma$. $\eta$ is related to the original $\alpha$, but not identical due to the spatial structure. The time-reversal-like symmetry ensures the exact degeneracy 
of the channels in the translationally invariant system. In a finite-size system with a defect, the degeneracy is approximate, but we can still identify the bulk channels, as the splitting is small and vanishes with increasing system size.

The localization, $f$-character, and pseudospin degeneracy of the half-filled defect states,
leads us to conclude that these states correspond to free moments (pseudospin doublets) bound to the defect center. At the level of the mean-field theory, these states do not interact with the heavy bulk states, but fluctuations lead to a two-channel Kondo interaction between the impurity moments and the heavy bulk states, and thus to MZMs.

\textbf{Origin of the Majorana: Gaussian fluctuations}

The localized states associated with Kondo hole and skyrmion defects are decoupled from the heavy conduction sea in the mean-field, but Gaussian fluctuations of the hybridization introduce a two-channel Kondo interaction between the defect and bulk states:
\begin{equation}\label{eq:2ckI}
    H_I=\sum_{\substack{\kappa, \kappa^{\prime}, \eta,\eta^{\prime}\\ \sigma, \sigma^{\prime}, \tau, \tau^{\prime}}}[J_I]_{\kappa,\kappa^{\prime}}^{\eta,\eta^{\prime}} \left(\beta^{\dagger}_{\kappa\eta \sigma}\vec{\sigma}_{\sigma,\sigma'}\beta_{\kappa^{\prime}\eta' \sigma'}\right)\cdot\left(\beta^{\dagger}_{I\tau}\vec{\sigma}_{\tau,\tau'}\beta_{I\tau'}\right).
\end{equation}
Here, $\eta$ labels the two distinct heavy conduction channels,
and this interaction is the most general possible. Note that channel degeneracy, $\left[J_I\right]^{\eta,\eta'} = J_I\delta_{\eta,\eta'}$ is not automatic
and multiple angular momentum screening channels are included via the $\kappa$ dependence, although we find that the usual s-wave channel dominates. 

The Kondo interaction arises from the interaction of our impurity and heavy bulk states with hybridization fluctuations, via tree-level diagrams of the form,
\begin{center}
\begin{fmffile}{diagram}
\begin{fmfgraph*}(140,60)
    \fmfleft{i1,o1}
    \fmfright{i2,o2}
    \fmf{fermion,label=$\kappa^{\prime} \eta^{\prime}\sigma^{\prime}$}{i1,v1}
    \fmf{scalar,label=$I\sigma^{\prime}$}{v1,o1}
    \fmf{fermion,label=$\kappa \eta\sigma$}{v2,o2}
    \fmf{scalar,label=$I\sigma$}{i2,v2}
    \fmf{dbl_wiggly}{v1,v2}
\end{fmfgraph*}
\end{fmffile}
\end{center}
Dashed lines indicate the impurity states, $\beta_{I \tau}$, and solid lines the bulk heavy fermions, $\beta_{\kappa\eta\sigma}$, while the squiggly line is the RPA propagator for hybridization fluctuations, obtained in the large-$N$ limit.  

We consider fluctuations of the hybridization, $\delta \mathbf{V}_j$ and constraint fields, $\delta \lambda_j$, which couple to the original fermions and give a correction to the Hamiltonian,
\begin{align}
    \delta H= &\sum_{j,\alpha, \sigma}\left(\delta V_{j\alpha} c_{j\alpha \sigma}^\dagger f_{j\sigma}+H.c.\right)\cr
   & +\sum_{j,\sigma}\delta \lambda_j f_{j\sigma}^{\dagger} f_{j\sigma} + \sum_{j,\alpha}\frac{2N |\delta V_{j\alpha}|^2}{J_K}.
\end{align}
The hybridization fluctuations can be decomposed into amplitude, $\delta|V|_j$ and angular, $\delta \theta_j$ and $\delta \phi_j$, fluctuations. The low energy angular fluctuations are hastatic Goldstone modes, which couple weakly to the emergent fermions, however, both amplitude and higher energy angular fluctuations contribute to the overall $J_I$.  

\begin{figure}
\includegraphics[width=0.95\columnwidth]{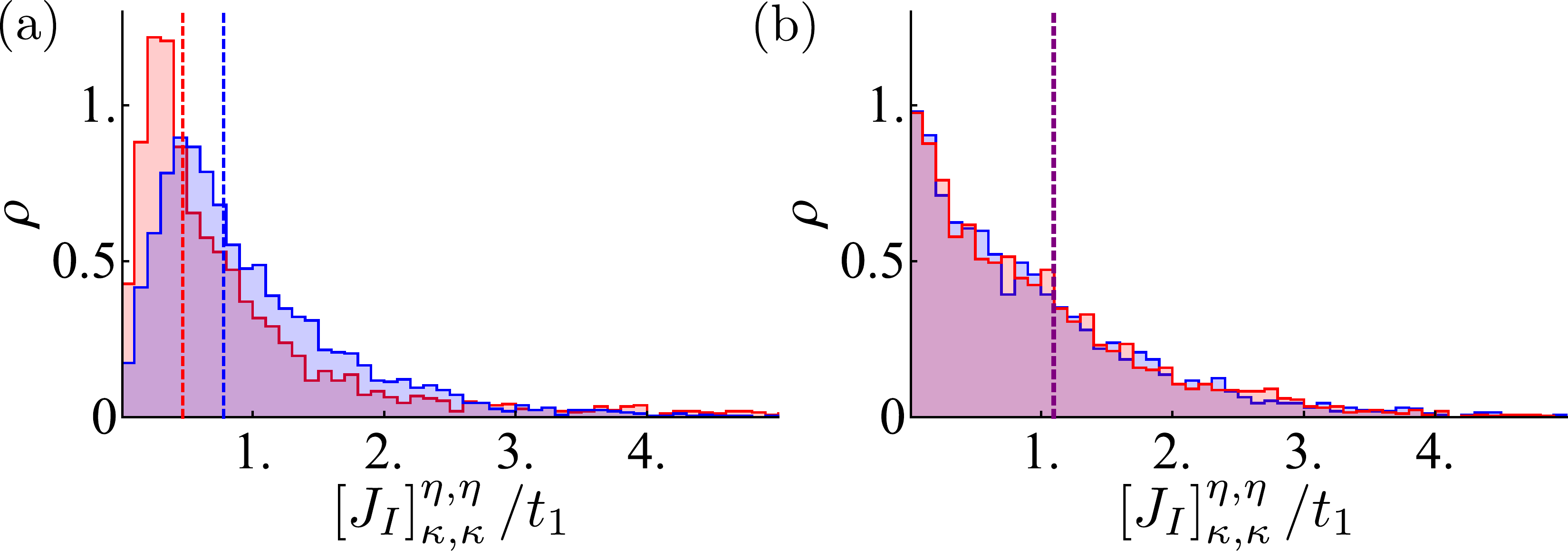}
\caption{\textbf{Distribution of impurity Kondo couplings} Histograms showing the distribution, $\rho$ of diagonal impurity Kondo couplings $\left[J_{I}\right]{}_{\kappa,\kappa}^{\eta,\eta}/t_1$ for the two Kondo channels (blue and red) in the cases of (a) hole and (b) skyrmion defects. The vertical lines show the leading $s$-wave coupling, $J_{I}^{(\eta)}$, obtained by averaging the distribution. $J_{I}^{(1)}$ is shown in blue and $J_{I}^{(2)}$ in red. For the skyrmions, the the two vertical lines overlap, while the Kondo hole has substantial asymmetry. This leads to the two white vertical lines drawn in the phase diagram, Fig.~\ref{fig:Fig1}. The relevant plot parameters are discussed in Methods. \label{fig:Fig4}
}
\vspace{-0.cm}
\end{figure}

Due to the broken translation symmetry, the calculation of the above diagram must be done numerically in real space, but is otherwise straightforward, with the details given in the Methods. 
Our calculation gives the full set of impurity Kondo interactions, $[J_I]_{\kappa,\kappa^{\prime}}^{\eta,\eta^{\prime}}$. We focus on components diagonal in $\kappa, \kappa'$, which are uniquely defined,
and the ones relevant for calculations like $T_I$  \cite{HewsonBook}; these are automatically diagonal in $\eta$.  The distribution of the diagonal Kondo couplings is shown in Fig.~\ref{fig:Fig4}, for both the skyrmion and Kondo hole cases. In what follows, we consider only the dominant $s$-wave contribution, $J_I^{(\eta)}$ obtained by averaging these distributions,
which leads to an effective two-channel Kondo Hamiltonian with a channel asymmetry, $\delta J_I=|J_I^{(1)}-J_I^{(2)}|/(J_I^{(1)}+J_I^{(2)})$ due to local symmetry breaking at the defect.  For the skyrmion, this symmetry breaking is weak, leading to asymmetries $\delta J_I\lesssim 1\%$. [see Fig.~\ref{fig:Fig4}(a) and Supplementary sections \ref{GF_Kondo} and \ref{sec:tfres}].  The Kondo hole breaks channel symmetry more strongly on-site, and has correspondingly larger asymmetries, $\delta J_I \sim 25\%$.  
While neither defect has perfect channel symmetry, we expect a wide range of two-channel Kondo physics for the skyrmion with its weak asymmetry. The range is bounded by $T_I > T>T_x$~\cite{rozhkov1998impurity, Cox1998, LeHur2016}. $T_I$ is the impurity Kondo temperature, and $T_x \sim \delta J_I^2 J_I^2/T_I$ is the crossover scale below which the impurity will be fully screened by the stronger channel~\cite{Pang1991}, estimated in Supplementary section~\ref{sec:asymmetry}, and schematically represented in Fig.~\ref{fig:Fig2}.

The Kondo insulating gap requires a sufficiently strong $J_I$ for a local moment to undergo the two-channel Kondo effect, which is satisfied when the impurity Kondo temperature, $T_I$ is greater than the gap, $\Delta/2$~\cite{coleman2015book,Balatsky2006, Komijani2020}.
Otherwise, the local moment remains unscreened.  While $T_I/\Delta$ depends on the microscopic details, we can estimate $T_I/T_K$. $T_K = \rho^{-1} \exp(-1/(2\rho J_K)$ is the single impurity Kondo temperature for the original lattice Kondo coupling, $J_K$, assuming a flat conduction electron density of states, $\rho$.  The impurity Kondo temperature is similarly $T_I = (\rho^*)^{-1}\exp(-1/(2\rho^* J_I)$, where $\rho^*$ is the average heavy density of states. While there is a hybridization gap at the Fermi energy, $\Delta \sim T_K$,
the spectral weight from the gap is redistributed to the side peaks, as shown in Fig. \ref{fig:Fig2}(d), with  average $\rho^*\sim T_K^{-1}$. The impurity Kondo temperature is then,
\begin{equation}
    T_I \approx T_K (\rho T_K)^{\frac{\rho T_K}{J_I/J_K}}.
\end{equation}
We find $J_I/J_K \gtrsim 1$ for all cases considered. For $\rho T_K = .01$, $T_I \approx .92 - .95 T_K$, depending on the defect type.  As the gap, $\Delta/2$ is typically smaller than $T_K/2$, we are safely within the impurity Kondo regime. A more careful treatment in Supplementary section \ref{sec:TIsizd} 
confirms the result.
Two-channel Kondo critical behavior will therefore be present for $T_x < T < T_I$ ~\cite{Cox1998, Zarand2000}, meaning that the impurity Kondo problem can be solved~\citep{AffleckLudwig_NPB_1991,Kivelson1992}, and the impurity state contains an emergent zero energy Majorana fermion~\cite{Kivelson1992,LeHur2013,LeHur2016,Folk2019} for $T_x<T<T_I$. 
For relevant ranges of $T_K/D$, $T_x \lesssim 0.005 T_I$ for $\delta J_I \lesssim 1\%$, and the skyrmion defect is \emph{generically} expected to have a wide region of Majorana physics. The Kondo hole case is more restrictive, as $T_x \lesssim T_I$ for $25\%$ asymmetry and the region of Majorana behavior is impractically small.

A Majorana bound state at zero energy is just the first requirement for Majorana zero modes with non-Abelian mutual statistics.  It must also be possible to adiabatically braid these defect states without exciting the system out of the ground state, protected by the Kondo insulating gap, $\Delta$. Moving the Kondo hole may be possible in artificial two-channel Kondo lattices, but is generally extremely disruptive, meaning Kondo holes are not good candidates for Majorana zero modes. For topological defects like our skyrmions, there is a well-defined notion of adiabatic transport and the braiding process can be performed smoothly, as long as the time scale associated with moving one skyrmion around the other is sufficiently large compared to the gap inverse.  
This braiding generically leads to non-commutative transformations within the ground-state manifold~\cite{dassarmareview}.

\textbf{Discussion}

We have shown that hybridization skyrmions in two-channel Kondo insulators can host Majorana zero modes with non-trivial mutual statistics, introducing Kondo materials as a new platform for topological quantum computation. Now we turn to possible experimental realizations and challenges, and discuss how defects might be detected and manipulated.

What is required is a sufficiently two-dimensional two-channel Kondo lattice, with the conduction electrons at quarter-filling and a sufficiently strong Kondo coupling.  Here, antiferrohastatic order will form and open up a Kondo insulating gap.  There are no such currently known materials, but there are several promising directions.  Antiferrohastatic order has been proposed to explain the hidden order state in URu$_2$Si$_2$ \cite{MydoshReview,Chandra2013}, a tetragonal material with a low carrier density in the hidden order. It is possible that a doped or gated thin film of URu$_2$Si$_2$ could be brought to quarter-filling, or that quarter-filled analogs could be grown.
There are also cubic Pr-based materials with two-channel Kondo physics, like PrTi$_2$Al$_{20}$ \cite{matsunami2011, sato2012}. These are relatively far from quarter-filling and form quadrupolar orders. If the filling could be appropriately tuned, these materials could be grown in a thin film geometry, and should form antiferrohastatic order as long as the local cubic symmetry breaking is sufficiently small. 
Alternatively, two-channel Kondo lattices can  be artificially engineered, either in quantum dot arrays~\cite{Oreg2007} or with non-Kramers Kondo ad-atoms, like Pr or U, on a quarter-filled metallic surface~\cite{Figgins2019}.

The most interesting complication is that realistic cases will typically open the Kondo insulating gap around a full Fermi surface, rather than the quadratic band touching considered here. It then becomes possible to get multiple doubly degenerate defect states in the gap, which, due to the strong impurity Kondo coupling, may act as a single defect state with a higher degeneracy. We consider the effect of a Fermi surface by changing conduction bandstructure parameters (see Supplementary sections \ref{supp:FS} and \ref{sec:tfres}), and we can find either one or two doubly degenerate states in the gap for the skyrmion defect, depending on parameters. The channel asymmetry of all the defect states is similar, meaning that a quasi-$4$-fold degenerate state should be treated as a two-channel $SU(4)$ impurity with some degree of $SU(4)$ symmetry breaking; the combined impurity is $3/4$ filled.  This case has not yet been well-studied, but is always \emph{fractionalized}, with either Majorana or more complicated parafermion statistics \cite{AliceaFendley_review_2016}.  

We considered a simple Kondo model, but real materials are described by a spin-orbit coupled Anderson model where the local symmetry affects the defect structure. In tetragonal symmetry, as relevant for URu$_2$Si$_2$, the $SO(3)$ symmetry of the composite order parameter is broken, and the spinor can either lie in the plane, with an effective $U(1)$ symmetry with weak $\mathbb{Z}_4$ pinning or out of the plane with $\mathbb{Z}_2$ symmetry \cite{Chandra2013,Kornjaca2020}.  Any candidate material would need to have the spinor in the plane, as proposed for hidden order in URu$_2$Si$_2$, and the stable topological defects are then hybridization vortices.  In the Supplementary sections \ref{def_types} and \ref{sec:tfres}, we examine a vortex defect state and find it to be comparable to the skyrmion defect states, with a nearly degenerate two-channel Kondo impurity interaction. The antiferrohastatic order in URu$_2$Si$_2$ is predicted to have staggered ferromagnetic layers \cite{Chandra2013}, and so the hybridization skyrmion could be induced by a capping material containing either ferromagnetic XY vortices or, more practically, ferromagnetic skyrmions.

In any candidate material, hastatic order could be confirmed by the combination of a Kondo insulating gap, a non-Kramers doublet ground state, and the signatures of broken translation and other symmetries \cite{Chandra2013,Zhang2018,Kornjaca2020}. Once the order is found, defect states must be generated and detected.  The most plausible route is to deposit a thin layer of an appropriate magnetic skyrmionic material on top.  Hybridization skyrmions couple to magnetic skyrmions with similar moment structure, and can be frozen in by cooling the combined system through the Kondo temperature. While antiferromagnetic skyrmion materials exist~\cite{Gao_2020_skyrmions}, it may be easiest to deposit the two-channel Kondo insulating material such that the terminating layer has a ferromagnetic structure and the antiferrohastatic skyrmions can be induced by ferromagnetic skyrmions in the capping material. The existence of the defect states can be straightforwardly confirmed by scanning tunneling microscopy, as the defect states are highly localized, with a localization length unrelated to the skyrmion radius, and are predominantly $f$ in character, which gives a strong Fano asymmetry~\cite{Maltseva2009}.
Controlling the skyrmion in the capping layer~\cite{Back2020} then controls the hybridization skyrmion, and can be used to braid these defects and execute quantum gates.

Two-channel Kondo systems provide an exciting new platform for realizing Ising anyons with Majorana zero modes. These relatively simple systems, however, may just be the tip of the iceberg, as we showed that some cases may instead host parafermions, when the effective defect degeneracy is greater than two.  Another route is to engineer $M$-channel Kondo impurities, which generically have a finite zero point entropy, $S = \ln 2 \cos \frac{\pi}{M+2}$ associated with a free anyon \cite{AffleckLudwig_NPB_1991,Sela2020}; in particular, \emph{three}-channel analogues of our hybridization skyrmions could realize Fibonacci anyons that can be used to construct all the necessary quantum gates for universal topological quantum computation~\cite{Nayak_RMP_2008}.


\textbf{Methods}

\footnotesize
\textbf{Diagonalization of mean-field defect Hamiltonian}

As the introduction of defects breaks translation symmetry, we examine the nature of the defect states by diagonalizing the mean-field Hamiltonian in real space.  To ensure the generality of our results, we examined a number of different defect types, parameter choices and lattice sizes.  Here, we detail our process, and give additional results in the Supplementary Information.

The real space diagonalization was performed for square-octagon lattices of $L \times L$ unit cells, where $L=7-21$ is odd in order to have a well defined center cell.  Each site can have two conduction electrons in the two channels ($\alpha$) and one $f$ electron, per pseudospin, for a total of $24 L^2$ degrees of freedom.

For all defect types, we mainly considered periodic boundary conditions to avoid surface states.  For the Kondo hole, Fig.~\ref{fig:Fig4}(b), an $A$ site is removed from the center unit cell with local chemical potential $\lambda_0$ on the site set to infinity and local hybridization $|V|_0 = 0$. The skyrmion defect has varying angles, $\theta$ and $\phi$, of the hastatic spinor, which now depend on the unit cell, $\mathbf{r}=(r_x,r_y)$,
\begin{align}\label{eq:skyrpatt}
\phi_{\bf r} =\arctan\frac{r_y}{r_x}+\phi_{0}, \quad \theta_{\bf r} =2\arctan\left(\frac{\sqrt{r_x^{2}+r_y^{2}}}{R_0}\right).
\end{align}
The angle $\phi_{0}$ allows to tune between different skyrmion types, with $\phi_{0}=0$ corresponding to hedgehog skyrmions, while $\phi_{0}=\pi/2$ corresponds to the N\'{e}el skyrmions. The skyrmion has a fixed core size, $R_0$, which we typically fix $R_0=5$ independent of the lattice size; similar results are found for $R_0 =2-5$ and all values of $\phi_0$.
In the main text, these are the angles of the hastatic spinor on the A site, with the other spinors in the same unit cell derived from Eq.~\eqref{eq:4sl}, which is the relevant defect if the intra-unit-cell AFH stiffness is larger than the inter-unit-cell stiffness.  Alternately, we can consider $\bf{r}$ to label the geometric position of each atom in Eq.~\eqref{eq:skyrpatt}, rather than unit cell coordinates; we call the first case a type I skyrmion and the second case a type II skyrmion.  It is possible to tune between the two types of skyrmions by varying the distance of atoms from the center of the square-octagon unit cell.  The defect physics does not qualitatively depend on this difference, as is shown in the Supplementary Information.

In tetragonal systems, hybridization vortices are the stable topological defects.  They may be obtained via Eq. (\ref{eq:skyrpatt}) with $\bf{r}$ by taking $\theta_{\bf r}=\pi/2$ and setting  $\phi_0=\pi/2$, with $\bf{r}$ the geometric position of each site; there is no core to these vortices, which are centered at the center of a square. 

The relevant Hamiltonian matrices are diagonalized numerically, and defect states identified using a modified version of the inverse-participation ratio, 
\begin{equation}
g_{\text{IPR}}\left[\psi\right]=\frac{1}{\sum_{\left|\mathbf{r}\right|<R_{c},a=A,B,C,D}\left|\psi_{f}\left(\mathbf{r},a\right)\right|^{4}}.
\end{equation}
Here, $\psi$ is the state eigenvector, $\mathbf{r}$ labels the unit cell and $a$ the site within the unit cell. The sum runs over only the $f$ degrees of freedom and is restricted to the core of the defect in real space by imposing $|\mathbf{r}| <R_{c}=2$ cutoff; note, there are no pure conduction electron defect states. The in-gap defect states minimize this quantity; defect states can be found both in the gap at the Fermi energy and in higher energy gaps.

In the main text, we choose $t_2 = 2t_1$ and quarter filling, but the ratio of $t_2/t_1$ can be varied substantially while maintaining the quadratic band touching at quarter-filling, with no qualitative effects.  The emergent $f$-hoppings, $\chi_1$ and $\chi_2$ are chosen to be small compared to $t_1$. We also have the mean-field parameters: the average constraint $\lambda$ and hybridization strength, $|V|$, as well as the chemical potential of the conduction electrons, $\mu$.  $\lambda$ is determined self-consistently in the translationally invariant case, and forces the defect state to be at zero energy, and thus half-filled.  $|V|$ must be sufficiently large compared to $\chi_1,\chi_2$ for a full Kondo insulating gap to open; for the main text, we choose $|V| = 0.5$, $\chi_1 = .01$, $\chi_2 = .04$, in units of $t_1$.  This choice of $|V|$ corresponds to some value of $J_K$, as discussed in the Supplementary Information. 

The real space diagonalization leads to the free fermionic modes labeled as $\beta_{\kappa\eta\sigma}$ in Eq.~\eqref{eq:2ckI}. These states can be related back to the original fermions by the unitary matrices, $U_{(i\alpha),(\kappa\eta)}$:
\begin{align}\label{eq:cftransf}
c_{i\alpha\sigma} =\sum_{\kappa\eta} U_{(i\alpha)(\kappa\eta)}\beta_{\kappa\eta\sigma}, \quad f_{i\sigma} =\sum_{\kappa\eta}U_{(i3)(\kappa\eta)}\beta_{\kappa\eta\sigma}.
\end{align}
$\alpha =1,2$ are the two channels of conduction electrons, while $\alpha =3$ denotes the f-electrons for convenience. $\kappa \eta \sigma$ here includes also the impurity states, $I\sigma$.  We use these matrix elements in the next section to derive the two-channel impurity Kondo coupling.

\textbf{Emergent impurity Kondo coupling}

To calculate the emergent impurity Kondo coupling via the tree-level diagram in the main text below Eq.~\eqref{eq:2ckI}, we must calculate both the fluctuation propagator in the RPA approximation, as well as the vertices that connect the emergent heavy bulk and impurity fermions to the fluctuations.  Working in the radial gauge, there are four fluctuation fields on each site: amplitude, $\delta |V|$; angular, $\delta \theta$ and $\delta \phi$; and constraint, $\delta \lambda$. As we are interested in defect states, we consider these fluctuations in real space.

The free propagator for the fluctuations is given by,
\begin{eqnarray}
\left[\chi_{0}\right]_{(i\mu)}^{(j\nu)}&=&\parbox[c][10pt][c]{130pt}{\centering
\begin{fmffile}{md1}
\begin{fmfgraph*}(80,10)
\fmfleft{i}
\fmfright{o}
\fmf{boson,tension=0.5}{i,o}
\fmflabel{$(i\mu)$}{i}
\fmflabel{$(j\nu)$}{o}
\end{fmfgraph*}
\end{fmffile}}
\end{eqnarray}
 \begin{equation*}
\left[\chi_{0}^{-1}\right]_{(i\mu)}^{(j\nu)}=N\delta_{ij}\left(\begin{array}{cccc}
\frac{2}{J_{K}}&0&0&0\\
0&0&0&0\\
0&0&0&0\\
0&0&0&0
\end{array}\right)_{\mu,\nu},
\end{equation*}
where $\mu,\nu=(\delta |V|,|V| \delta \theta, |V| \delta \phi, \delta \lambda)$ labels the fluctuation fields, and we include $|V|$ in the angular fields to keep the units consistent. $i, j$ denote the sites (here, this consists of both the unit cell, $\mathbf{r}$ and $a = A,B,C,D$, but we use this shorthand for convenience). 

While conventional calculations of fluctuation corrections work with the original $c$ and $f$ fermions, here we find it more convenient to work with the emergent heavy fermions, where the fluctuations couple to the heavy conduction electron states ($\kappa\eta$) and impurity states via complicated vertices.  

As the original couplings are simplest in Cartesian gauge, it is convenient to relate the real fluctuation fields in radial gauge to the complex $(\delta V_1,\delta V_2)$, with:
\begin{equation}
    \begin{pmatrix}
\delta |V|_i \\
|V| \delta \theta_i \\
|V| \delta \phi_i\\
\delta \lambda_i
\end{pmatrix}=
    \begin{pmatrix}
\frac{\partial V_{i1}}{\partial |V|} &\frac{\partial V_{i2}}{\partial |V|} & 0\\
\frac{1}{|V|}\frac{\partial V_{i1}}{\partial \theta} &\frac{1}{|V|}\frac{\partial V_{i2}}{\partial \theta} & 0\\
\frac{1}{|V|}\frac{\partial V_{i1}}{\partial \phi} &\frac{1}{|V|}\frac{\partial V_{i2}}{\partial \phi} & 0\\
0 &0 & \frac{1}{2}
\end{pmatrix}
    \begin{pmatrix}
\delta V_{i1} \\
\delta V_{i2} \\
\delta \lambda_i
\end{pmatrix},
\end{equation}
and similarly for conjugate fields.  $\delta \lambda_i$ is, of course, always a real field, and the factor of $1/2$ is necessary only to package it together with the complex $\delta V_{i\alpha}$.  We call the site-dependent matrix above $s_{\mu\xi}^i$, with $\xi = (\delta V_1,\delta V_2, \delta \lambda)$. 
The vertices are then:

\begin{eqnarray}
\parbox[c][60pt][c]{60pt}{\centering
\begin{fmffile}{md2}
\begin{fmfgraph*}(60,60)
    \fmfleft{i1}
    \fmflabel{$(i\mu)$}{i1}
    \fmfright{o1,o2}
    \fmf{boson,tension=1.}{i1,w1}
    \fmf{fermion,label=$\kappa^{\prime}\eta^{\prime}\sigma$,label.side=left}{o1,w1}
    \fmf{fermion,label=$\kappa\eta\sigma$,label.side=left}{w1,o2}
    \fmfdot{w1}
\end{fmfgraph*}
\end{fmffile}}&=&W_{(\kappa\eta),(\kappa^{\prime}\eta^{\prime})}^{(i\mu)}+W_{(\kappa^{\prime}\eta^{\prime}),(\kappa\eta)}^{*(i\mu)}
\end{eqnarray}
where
\begin{equation}
    W_{(\kappa\eta),(\kappa^{\prime}\eta^{\prime})}^{(i\mu)}=\sum_{\xi}s^i_{\mu\xi}U^*_{(i\xi),(\kappa\eta)}U_{(i3),(\kappa^{\prime}\eta^{\prime})}
\end{equation}
is independent of $\sigma$.
$U$ is the unitary transformation matrix defined in Eq. (\ref{eq:cftransf}) in the real-space diagonalization.

We now calculate the full fluctuation propagator in the RPA approximation, which is justified by the large-$N$ approximation up to $\mathcal O(1/N)$~\cite{coleman2015book}.  We also limit ourselves to zero frequency, as the tree level diagrams give the leading non-trivial contribution. The RPA propagator,
\begin{eqnarray}
\left[\chi_{RPA}\right]_{(i\mu)}^{(j\nu)}&=&\parbox[c][10pt][c]{130pt}{\centering
\begin{fmffile}{md4}
\begin{fmfgraph*}(80,10)
\fmfleft{i}
\fmfright{o}
\fmf{dbl_wiggly,tension=0.5}{i,o}
\fmflabel{$(i\mu)$}{i}
\fmflabel{$(j\nu)$}{o}
\end{fmfgraph*}
\end{fmffile}}
\end{eqnarray}
is given in matrix form by
\begin{equation}
    \chi_{RPA}^{-1}=-\chi_0^{-1}+\Pi,
\end{equation}
with $\Pi$ denoting the polarization bubble:
\begin{eqnarray}\label{eq:polbub}
\Pi_{(i\mu)}^{(j\nu)}&=&\parbox[c][60pt][c]{130pt}{\centering
\begin{fmffile}{md3}
\begin{fmfgraph*}(110,60)
\fmfleft{i}
\fmfright{o}
\fmf{phantom}{i,v1}
\fmf{phantom}{v2,o}
\fmf{fermion,label=$\kappa\eta$,left=0.5,tension=0.1}{v1,v2}
\fmf{fermion,label=$\kappa^{\prime}\eta^{\prime}$,label.side=right,left=0.5,tension=0.1}{v2,v1}
\fmfdot{v1,v2}
\fmflabel{$(i\mu)$}{v1}
\fmflabel{$(j\nu)$}{v2}
\end{fmfgraph*}
\end{fmffile}}
\end{eqnarray}
\begin{align*}
    =&-N\sum_{\substack{(\kappa\eta),\\(\kappa^{\prime}\eta^{\prime})}}\left(W_{(\kappa\eta),(\kappa^{\prime}\eta^{\prime})}^{(i\mu)}+W_{(\kappa^{\prime}\eta^{\prime}),(\kappa\eta)}^{*(i\mu)}\right)\\
    &\left(W_{(\kappa^{\prime}\eta^{\prime}),(\kappa\eta)}^{(j\nu)}+W_{(\kappa\eta),(\kappa^{\prime}\eta^{\prime})}^{*(j\nu)}\right)\frac{n_f(E_{\kappa\eta})-n_f(E_{\kappa^{\prime}\eta^{\prime}})}{E_{\kappa\eta}-E_{\kappa^{\prime}\eta^{\prime}}}.
\end{align*}
We evaluate the Fermi functions at zero temperature.
We can now calculate the Kondo interaction to leading order, as captured by the following tree-level diagram with the full propagator,
\begin{eqnarray*}
\parbox[c][65pt][c]{130pt}{\centering
\begin{fmffile}{md5}
\begin{fmfgraph*}(120,60)
    \fmfleft{i1,o1}
    \fmfright{i2,o2}
    \fmfv{label=$(i\mu)$,label.angle=-68}{v1}
    \fmfv{label=$(j\nu)$,label.angle=-112}{v2}
    \fmf{fermion,label=$\kappa^{\prime} \eta^{\prime} \tau$}{i1,v1}
    \fmf{scalar,label=$I\tau$}{v1,o1}
    \fmf{fermion,label=$\kappa \eta\sigma$}{v2,o2}
    \fmf{scalar,label=$I\sigma$}{i2,v2}
    \fmf{dbl_wiggly}{v1,v2}
    \fmfdot{v1,v2}
\end{fmfgraph*}
\end{fmffile}}&=&T_{(\kappa\eta)}^{(\kappa^{\prime}\eta^{\prime})}
\end{eqnarray*}
Here, the dashed lines indicate the impurity states, while the solid lines indicate the heavy bulk states. The diagram represents spin-independent interaction terms that lead to a Kondo interaction, as well as a potential scattering term that we neglect. The interaction can be rewritten using the $SU(2)$ Pauli matrix completeness relation, $2\delta_{\sigma \tau}\delta_{\lambda \rho} = \delta_{\sigma\rho}\delta_{\lambda \tau }+\vec{\sigma}_{\sigma \rho} \cdot \vec{\sigma}_{\lambda \tau}$:  
\begin{equation}
    \beta\dg_{\kappa\eta\sigma}\beta_{I\sigma}\beta\dg_{I \tau}\beta_{\kappa'\eta'\tau} = -\frac{1}{2}\left(\beta^{\dagger}_{\kappa\eta \sigma}\vec{\sigma}_{\sigma,\sigma'}\beta_{\kappa^{\prime}\eta' \sigma'}\right)\cdot\left(\beta^{\dagger}_{I\tau}\vec{\sigma}_{\tau,\tau'}\beta_{I\tau'}\right).
\end{equation}
Here, we use Einstein summation notation for convenience. 

The diagram is mathematically expressed as,
\begin{align}
\label{eq:treeval}
    T_{(\kappa\eta)}^{(\kappa^{\prime}\eta^{\prime})}&=\sum_{(i\mu),(j\nu)}\left(W_{(I),(\kappa^{\prime}\eta^{\prime})}^{(i\mu)}+W_{(\kappa^{\prime}\eta^{\prime}),(I)}^{*(i\mu)}\right)\\
    &\left(W_{(\kappa\eta),(I)}^{(j\nu)}+W_{(I),(\kappa\eta)}^{*(j\nu)}\right)\left[\chi_{RPA}\right]_{(i\mu)}^{(j\nu)}\notag
\end{align}
which gives the Kondo interaction:
\begin{align}
    J_{I(\kappa,\kappa^{\prime})}^{(\eta,\eta^{\prime})}=- T_{(\kappa\eta)}^{(\kappa^{\prime}\eta^{\prime})},
\end{align}
with the internal sum over all sites, $i,j$, as well as the four fluctuation channels, $\mu, \nu$.  These sums can be evaluated numerically for any lattice, defect type, and set of parameters. 

While the resulting Kondo interaction is a matrix in $\kappa,\kappa'$, we can focus on the analysis of the diagonal Kondo couplings, $J_{I(\kappa,\kappa)}^{(\eta)}$, as these are relevant for calculations \cite{HewsonBook}. The terms diagonal in $\kappa$ terms are automatically diagonal in the channel, $\eta = \eta'$, although the two channels can have different couplings for a given $\kappa$. These diagonal terms are always antiferromagnetic, as the vertices in Eq.~\eqref{eq:treeval} can be written as vectors, $v_{\kappa\eta}^{(i\mu)}$, writing
\begin{align}
    J_{I(\kappa,\kappa)}^{(\eta)}= v_{\kappa\eta}^{(i\mu)} \left[-\chi_{RPA}\right]_{(i\mu)}^{(j\nu)} \left[v_{\kappa\eta}^{(j\nu)}\right]^\dagger.
\end{align}
Here, we can rewrite this as $\langle v|-\chi_{RPA}|v\rangle = \sum_\lambda |\langle v|\lambda\rangle|^2 \omega_\lambda$, where $\omega_\lambda$ and $|\lambda\rangle$ are the eigenvalues and eigenvectors of $-\chi_{RPA}$.  If hastatic order is stable, all eigenvalues of $-\chi_{RPA}$ are positive, which is also what we find numerically, with the exception of the vortex state, as discussed in the Supplementary section  \ref{GF_prop}.
The distribution of these coupling strengths is shown in Fig.~\ref{fig:Fig4}. In the Supplementary sections \ref{GF_Kondo} and \ref{sec:tfres}, we explore how these distributions vary with the nature of the defect, system size, and $J_K$.

These Kondo couplings are predominantly s-wave, $J_I^{(\eta)} = \sum_\kappa J_{\kappa,\kappa}^{(\eta)}$,  as this channel dominates over any higher angular momentum channels; we can see this by calculating $\frac{1}{L^2}\sum_{\kappa,\kappa'}|J_{\kappa,\kappa'}^{\eta,\eta}| \approx J_I^{(\eta)}$.


\textbf{Acknowledgements} We acknowledge valuable discussions about Majorana fermions with T. Iadecola and the experimental feasibility of our proposal with S. Paschen. We also thank J. van Dyke for discussion in the early stages of the project. M.K., V.L.Q, and R.F acknowledge financial support from the U.S. Department of Energy, Office of Science, Basic Energy Sciences, under Award DE-SC0015891.


 

\bibliographystyle{apsrev4-2}
\bibliography{hastatic_defects}

\clearpage
\normalsize
\onecolumngrid

\title{Supplementary Information}

The Supplementary Information is organized as follows. In Section~\ref{square-octagon}, we write down the square-octagon model without defects. In Section~\ref{topo-defects}, we discuss the topologically stable defects in antiferrohastatic orders. In Section~\ref{real-space-diagon}, we show details of the real-space numerical diagonalization in the presence of defects, including the spectrum and the wave functions for different defect types. Gaussian fluctuations are explored, starting with the RPA propagator, in Section~\ref{GF_prop}. In Section~\ref{GF_Kondo}, we show estimates for the effective impurity Kondo coupling and the channel asymmetry. We finish by showing the full range of impurity Kondo coupling results for different defect types and parameters, in Section~\ref{sec:tfres}.

\section{The square-octagon model in momentum space~\label{square-octagon} }

The square octagon lattice was shown in Fig.~2(a) of the main text, where the key feature is that there are four sites per unit cell that we label $a =A,B,C,D$.  Here, we start with the mean-field decoupled model that is quadratic in both conduction electrons and the auxiliary fermions representing the local moments. We generically consider conduction electrons ($c$) and auxiliary fermions ($f$) hopping on this lattice with $t_{ij}^{a a'}$and $\chi_{ij}^{a a'}$, respectively.  $\chi$ are emergent $f-f$ hopping terms that we treat as free parameters, although they can be derived from Heisenberg spin interactions of the local moments \cite{Andrei1989}. 

The generic conduction electron term is written,
\begin{equation}
H_{c}=-\sum_{i,j, a, a',\alpha,\sigma}t_{i,j}^{a a'}c_{ia\alpha\sigma}^{\dagger}c_{ja'\alpha\sigma}-\mu\sum_{i,\sigma,\alpha}c_{ia\alpha\sigma}^{\dagger}c_{ia\alpha\sigma},
\end{equation}
where $\sigma$ is the pseudo-spin and $\alpha$ is the channel index. For most of this work we only consider nearest neighbor ($t_1$) and next-nearest neighbor ($t_2$) hoppings, that are within and between unit cells, respectively.  We consider fourth neighbor hoppings ($t_4$) in Section \ref{supp:FS}, when we examine a model with a Fermi surface at quarter-filling rather than a quadratic band touching.  The chemical potential, $\mu$ is tuned self-consistently to keep the conduction electrons at quarter-filling. This term is diagonal in both pseudo-spin and channel space. The conduction electron Hamiltonian in momentum space is then,
\begin{equation}
H_{c}\left(\boldsymbol{k}\right)=-\left(\begin{array}{cccc}
\mu & t_{1} & t_{2}e^{-ik_{x}} & t_{1}\\
t_{1} & \mu & t_{1} & e^{-ik_{y}}t_{2}\\
t_{2}e^{ik_{x}} & t_{1} & \mu & t_{1}\\
t_{1} & t_{2}e^{ik_{y}} & t_{1} & \mu
\end{array}\right),
\end{equation}
written in the $A,B,C,D$ basis for a single pseudo-spin and channel species. For the special
case of $t_{2}/t_{1}=1$, three bands cross at the $\Gamma$ point.
For $|t_2|>|t_1|$, one of the band separates from the other two, leaving a quadratic band touching at the $\Gamma$ point at 1/4 filling for $t_1, t_2 < 0$.  Exactly the same physics would be obtained with positive hoppings, but at 3/4 filling. In the main text, without loss of generality, we keep $t_{2}/t_{1}=2$. Small further neighbor hoppings do not qualitatively change the quadratic band touching.

The auxiliary fermion Hamiltonian is generically given by
\begin{equation}
H_{f}=\sum_{i,j,a,a',\sigma}\left(\chi_{ij}^{a a'}f_{ia\sigma}^{\dagger}f_{ja'\sigma}+\left(\chi_{ij}^{aa'}\right)^{*}f_{ja'\sigma}^{\dagger}f_{i a \sigma}\right)+\sum_{i,a,\sigma}\lambda_{ia}\left(f_{ia\sigma}^{\dagger}f_{ia\sigma}-1\right),
\end{equation}
where $\lambda_{ia}$ is a Lagrange multiplier enforcing the half-filling of the fermions on each site.  In the mean-field ansatz, we assume $\lambda_{ia} = \lambda$ is uniform and keep only real nearest ($\chi_1$) and next-nearest ($\chi_2$) hopping terms.  Again in the four-site basis, the Hamiltonian for a single pseudo-spin species is,
\begin{equation}
H_{f}\left(\boldsymbol{k}\right)=\left(\begin{array}{cccc}
\lambda & \chi_{1} & \chi_{2}e^{-ik_{x}} & \chi_{1}\\
\chi_{1} & \lambda & \chi_{1} & \chi_{2}e^{-ik_{y}}\\
\chi_{2}e^{ik_{x}} & \chi_{1} & \lambda & \chi_{1}\\
\chi_{1} & \chi_{2}e^{ik_{y}} & \chi_{1} & \lambda
\end{array}\right).
\end{equation}

Finally, we treat the on-site hybridization between $c$ and $f$, which is generically
\begin{equation}
H_{c,f}=\sum_{i,a,\sigma,\alpha}\left(V_{ia\alpha}c_{ia\alpha\sigma}^{\dagger}f_{ia\sigma}+H.c.\right).
\end{equation}
As the channel symmetry is important, we choose the four sublattice pattern given in the main text, $V_{j a \sigma} = \pm |V|\delta_{\sigma,+}$ for A, C sublattices and $V_{j a \sigma} = \pm |V|\delta_{\sigma,-}$ for B,D sublattices, with the spinors taken along $\pm \hat z$ without loss of generality due to the $SU(2)$ channel symmetry.  This term is still independent of the pseudo-spin, but not of the channel index and so we do not show the (diagonal) 8x8 matrix form.
The complete large-$N$ mean-field AFH Hamiltonian combines these terms as:
\begin{equation}
H=H_{c}+H_{f}+H_{c,f}+\sum_{ja}\frac{N |{\bf V}_{ja}|^2}{J_K}.\label{eq:full_MF_Hamilt}
\end{equation}

This model is then solved self-consistently for $|V|$, $\lambda$ and $\mu$, which gives the Kondo insulating gap at 1/4-filling~\cite{Zhang2018}.  Note that in more realistic Hamiltonians the spinor will feel the lattice symmetry, and thus will not have the full $SU(2)$ symmetry. There, the angles representing the spinor direction, $\theta$, $\phi$ should also be solved for self-consistently, and the spinor will be pinned along high symmetry directions~\cite{Chandra2013, Zhang2018}.

The exact value of $|V|$, which controls the gap magnitude can be tuned by changing $J_K$; for simplicity we work directly with $|V|$, which can be mapped back to $J_K$ when desired. If the $|\chi|$ are nonzero, a full insulating gap requires a finite $|V|$.  We solve for $\lambda$ and $\mu$ self-consistently by forcing the f fermions to be half-filled on average and the conduction electrons to be quarter filled. The corresponding hybridized band structure is shown in Fig.~\ref{fig:Fig2} of main text. As noted in the main text, the time-reversal-like symmetry preserved by the 4SL ansatz manifests via Kramers partners located at different points in momentum space.

\section{Topological defects in Hastatic order~\label{topo-defects}}

The stable topological defects for any order parameter symmetry in any dimension may be found by examining the homotopy groups of the order parameter manifold~\cite{Mermin1979}. In the large-$N$ limit, the symmetry group of hastatic order is $G = SU(2)$.  If the spinors form a collinear order, the $SU(2)$ symmetry is broken down to $H=U(1)$: the remaining degree of freedom perpendicular to the spinor direction.  The order parameter manifold is therefore $R = G/H = SU(2)/U(1) = S_2$, where $S_2$ is the two-dimensional surface of a sphere.  

In two spatial dimensions, the homotopy groups $\pi_{1}(R)=\pi_{1}(S_{2})=0$, indicating that vortices are not topologically stable.  However, $\pi_{2}(R)=\pi_{2}(S_{2})=\mathbb{Z}$ shows that 2D skyrmions are indeed topologically stable.

There are two possible complications to consider: (1) what happens to the order parameter symmetry for finite-$N$? and (2) what happens in cases where the hastatic spinor has a lower symmetry group due to coupling to the lattice?  For finite-$N$, we should instead consider the composite order parameter, $\mathbf{V}\dg\vec{\sigma}\mathbf{V}$, which is gauge invariant.  This order parameter symmetry is $G = SO(3)$, as for magnetic orders.  $H = U(1)$ for collinear orders, and $R = G/H = SO(3)/U(1) = S_2$, leaving the previous analysis unaffected.  Note that at half-filling, the composite order parameter is actually an $SO(5)$ order parameter, as there are additional degenerate composite pairing order parameters~\cite{Hoshino2011}, but we are always away from half-filling.  In lower symmetries, the symmetry group of the hastatic order is reduced from $SO(3)$ down to the point group of a crystal. However, the in-plane pinning in tetragonal symmetry and pinning in the cubic symmetry are extremely weak~\cite{Zhang2018, Chandra2013}. Therefore it is reasonable to consider $G=SO(3)$ and $G=SO(2)$ symmetries for the cubic and tetragonal/hexagonal in-plane moment symmetries, respectively. If the composite order parameter points along the $\hat z$ axis in tetragonal or hexagonal symmetry, $G = \mathbb{Z}_2$ and there are no stable point like defects.  For the in-plane order, $G = SO(2)$ and any ordering fully breaks the symmetry. $R = SO(2)$ and $\pi_{1}(R)=\pi_{1}(S_{1})=\mathbb{Z}$ gives XY vortices as the stable topological defects in 2D.

\section{Real-space diagonalization results~\label{real-space-diagon}}

The main text presented two real-space diagonalization results, for the Kondo hole and one type of skyrmion, for one set of parameters.  In this section, we give more details for these two cases and examine a wide variety of other defect types and parameters to understand the robustness of the defect behavior. In all cases, we numerically diagonalize the square-octagon lattice with $N_s = L\times L$ unit cells with periodic boundary conditions to avoid surface states.  In section \ref{def_types}, we look at the four defect types, while in section \ref{supp:scaling} we examine the dependence of our results on $L$, the skyrmion radius $R_0$ and the size of the insulating gap. In section \ref{supp:FS}, we examine a modified model where the conduction electrons have a Fermi surface rather than a quadratic band touching at quarter-filling.  We also examined the effect of varying the parameters ($t_1,t_2$, $\chi_1,\chi_2$), which we do not show here as they have no qualitative effects on the nature of the defect states.

\subsection{Defect types~\label{def_types}}

We considered four different defect types, as defined in the Methods. For each type, we plot an example energy spectrum, modified inverse participation ratio ($g_{IPR}$) and the spatial dependence of the $f$ part of the wave function of the defect state.

\begin{figure}
\includegraphics[width=0.95\columnwidth]{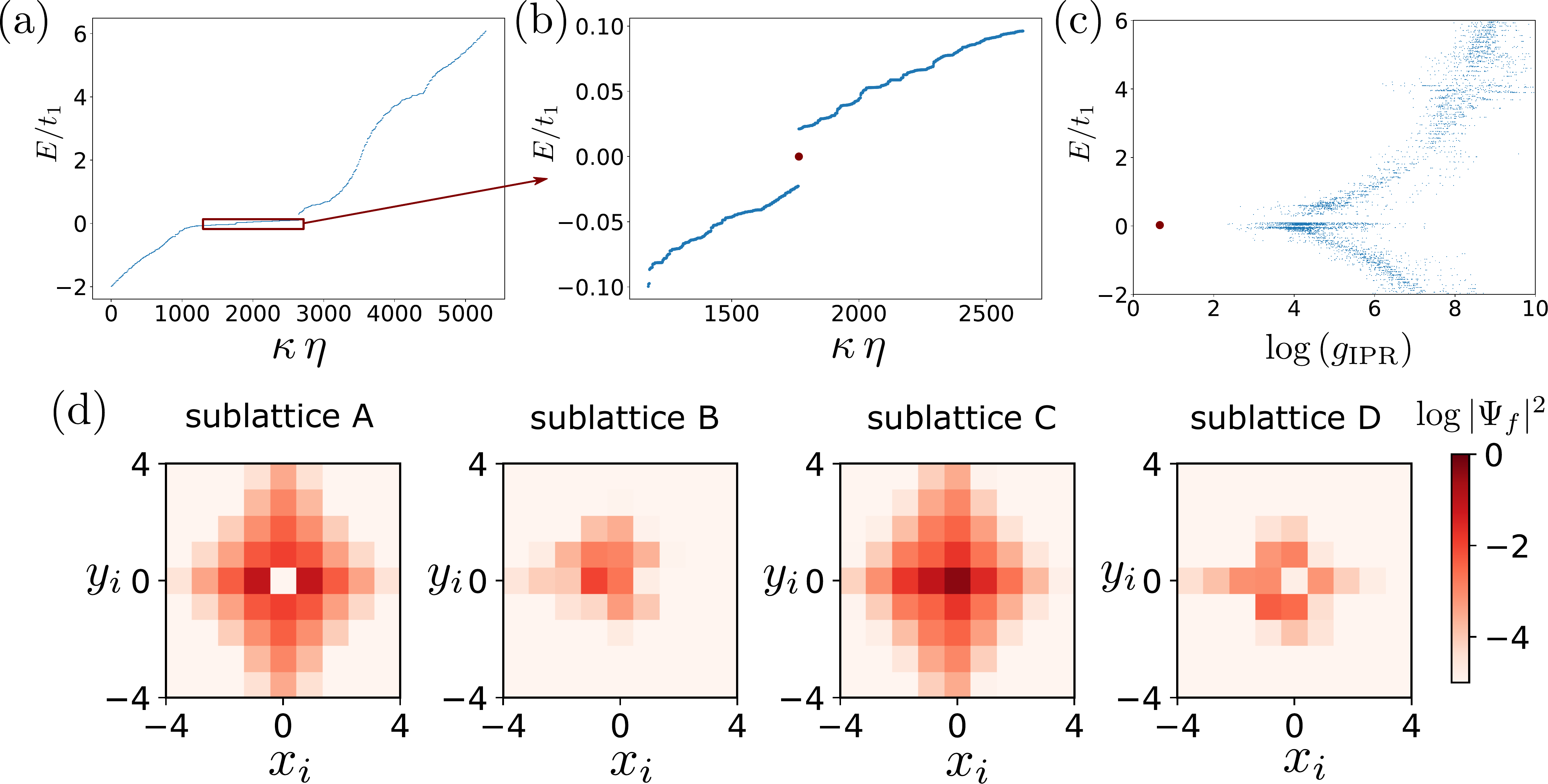}
\caption{Kondo hole defect state, with parameters $\mu=-2$, $t_{2}=-2$, $\chi_{1}=0.01$, $\chi_{2}=0.04$
and $|V|=0.5$. (a) Energy spectrum; (b) the region close to zero energy, showing a localized in-gap state (red dot). The $x$-axis is an arbitrary eigenstate label.  (c) shows the modified IPR for all eigenstates. The red dot indicates the localized defect state. In (d), the $f$-part of the localized defect state wave function in the real space on four sublattices. Since the Kondo hole corresponds to a missing magnetic moment at the center $A$ sublattice, the wave function has no amplitude at that site (first panel of (d)), while most of its weight is at the C sublattice in the same unit cell and a few surrounding unit cells. 
\label{fig:defectWF}}
\end{figure}

\begin{figure}
\includegraphics[width=0.95\columnwidth]{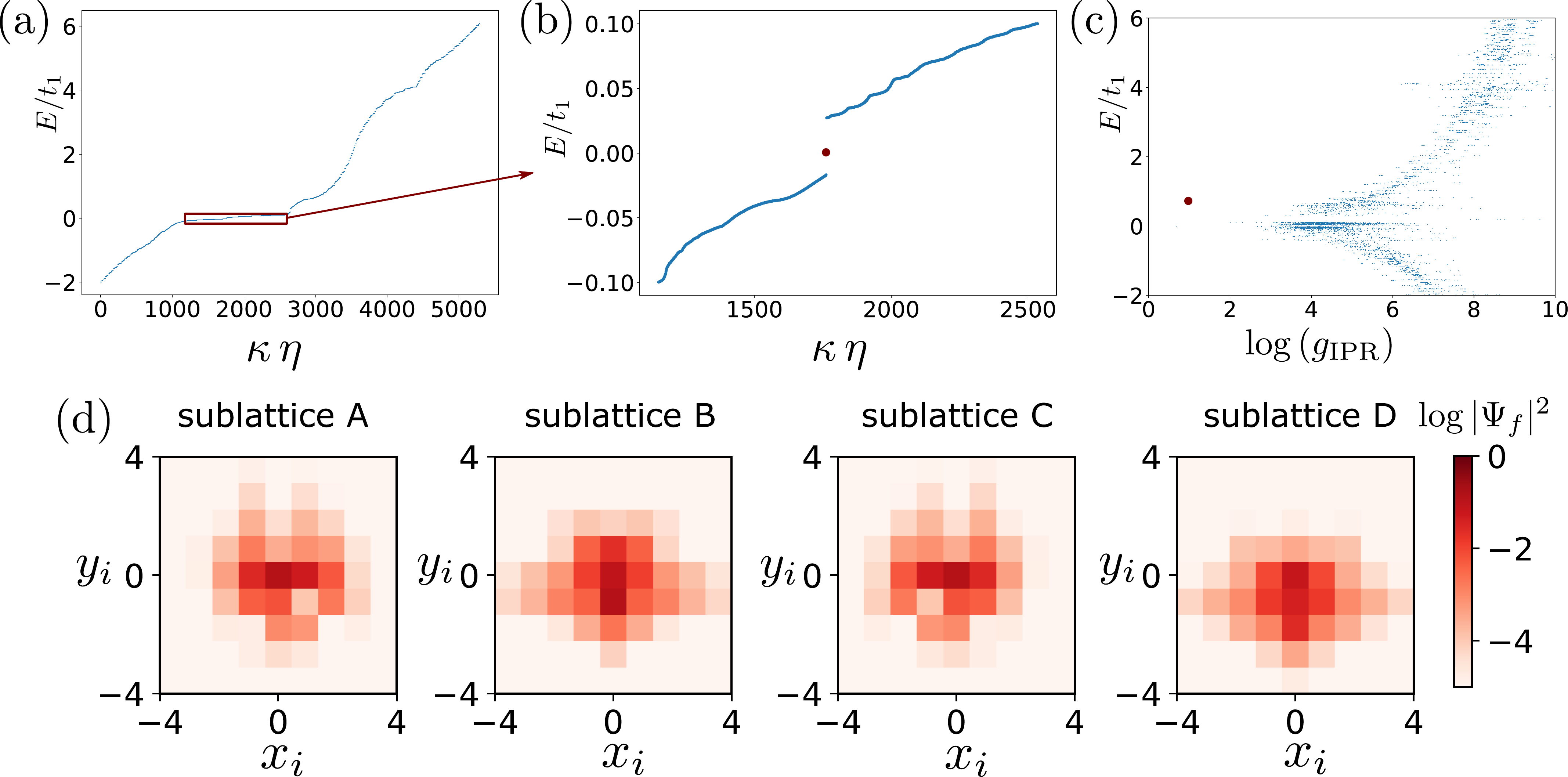}
\caption{Skyrmion defect state (type I) with parameters $\mu=-2$, $t_{2}=-2$, $\chi_{1}=0.01$, $\chi_{2}=0.04$
and $|V|=0.5$. Each panel is similar to the previous Kondo hole figure. (a) shows the energy spectrum, (b) a narrow region near zero energy showing the localized in-gap state with a red dot. The $x$-axis is an arbitrary eigenstate label. In (c), the modified IPR for all eigenstates is shown, with the red dot indicating the localized defect state. In (d), the $f$-part of the localized defect state wave function is shown in real space on the four sublattices. The wave function is not $C_{4}$ symmetric, as the skyrmion weakly breaks the square lattice symmetry. The wave function has an appreciable amplitude on all four sublattices, in contrast to the Kondo hole. \label{fig:skyrWF}}
\end{figure}

\begin{figure}
\includegraphics[width=0.95\columnwidth]{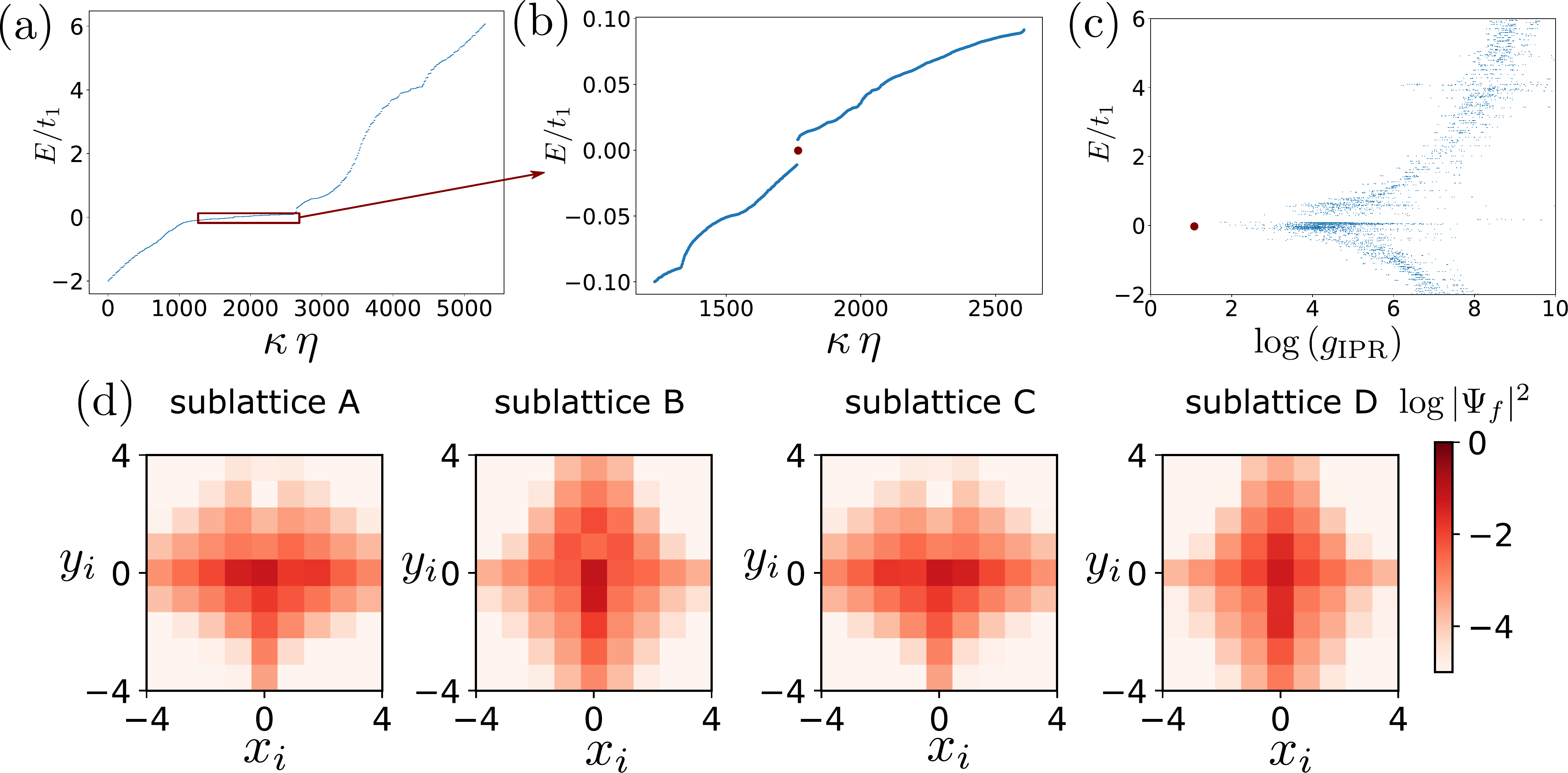}
\caption{Skyrmion defect state (type II) with parameters $\mu=-2$, $t_{2}=-2$, $\chi_{1}=0.02$, $\chi_{2}=0.04$ and $|V|=0.5$. Each panel is similar to the previous two figures. (a) shows the energy spectrum, (b) a narrow region near zero energy showing the localized in-gap state with a red dot. The $x$-axis is an arbitrary eigenstate label. In (c), the modified IPR for all eigenstates is shown, with the red dot indicating the localized defect state. In (d), the $f$-part of the localized defect state wave function is shown in real space on the four sublattices. Despite a somewhat different spinor arrangement, this defect state is very similar to the type I skyrmion defect. \label{fig:NskyrWF}}
\end{figure}

\begin{figure}
\includegraphics[width=0.95\columnwidth]{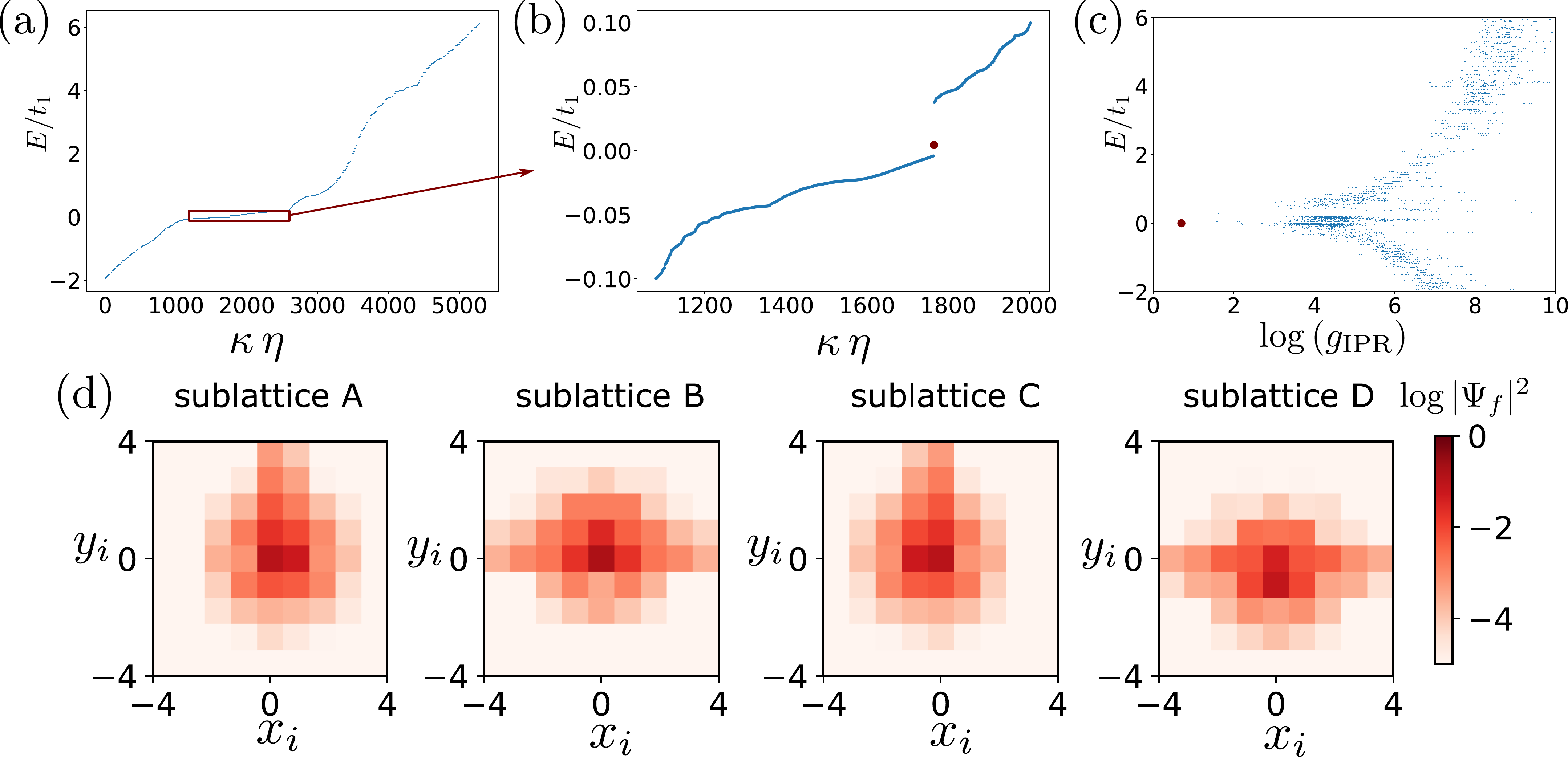}
\caption{Vortex defect state, with parameters $\mu=-2$, $t_{2}=-2$, $\chi_{1}=0.01$, $\chi_{2}=0.08$
and $|V|=0.5$. Each panel is similar to previous figures. (a) shows the energy spectrum, (b) a narrow region near zero energy showing the localized in-gap state with a red dot. The $x$-axis is an arbitrary eigenstate label. In (c), the modified IPR for all eigenstates is shown, with the red dot indicating the localized defect state. In (d), the $f$-part of the localized defect state wave function is shown in real space on the four sublattices.  Despite a different spinor arrangement, especially in the central cell, the defect state has localization and symmetry properties similar to both skyrmion states. \label{fig:vornWF}}
\end{figure}

The Kondo hole state is shown in Fig.~\ref{fig:defectWF} with $\mu=-2$, $t_{2}=-2$, $\chi_{1}=0.01$, $\chi_{2}=0.04$ and $|V|=0.5$, where $L = 21$. The localized state appears in the middle of the Kondo insulating gap, as seen in Fig.~\ref{fig:defectWF}(b). There is a single state per pseudo-spin. The modified IPR, defined in the Methods, is shown in Fig.~\ref{fig:defectWF}(c), and indicates that the state is localized on $\mathcal{O}(1)$ sites, compared to all other states with IPRs of the order of the system size. In Fig.~\ref{fig:defectWF}(d), we show that the defect state wave function is localized at the center of the lattice; the amplitude is given in log-scale, which makes the exponential nature of the localization clear. 
We use the same set of parameters for the type I skyrmion, where the spinors within a single unit cell are perfectly aligned, and the results are shown in Fig.~\ref{fig:skyrWF}. Again, $L =21$ and the skyrmion radius, $R_0 = 5$. The type II skyrmion involves a smooth variation of the spinor within and between unit cells; here, we increase $\chi_1 =0.02$. 
These results are shown in Fig. \ref{fig:NskyrWF}. Finally, we consider a vortex defect in Fig. \ref{fig:vornWF}, even though this is topologically unstable, where we increase $\chi_2 = .08$. 
In addition to the two types of skyrmions, it is also possible to tune the skyrmion from N\'{e}el to hedgehog by changing $\phi_0$; we use $\phi_0 = \pi/2$ for N\'{e}el type, but changing $\phi_0$ has very little effect.

While some details vary, the defect state is generically present, is always predominantly $f$ in character, is always doubly degenerate and is always localized, as long as there is a full gap. For the skyrmion and vortex cases, there is another requirement to get a localized state: the $\chi$'s connecting moments with non-trivial angles must be nonzero. $\chi_2$ must be non-zero for the type I skyrmion, and both $\chi_1$ and $\chi_2$ must be non-zero for the type II skyrmion and the vortex.  $\chi_2/\chi_1$ must also be sufficiently large, which is likely an artifact of the square-octagon lattice, where $\chi_1$ is not dispersive. 
The symmetry properties are significantly different between the Kondo hole and the smoother topological defects.  The Kondo hole defect state is orthogonal to the hole location (on the center $A$ site), with most weight on the center $C$ site, and almost no weight on the $B$ and $D$ sublattices. The skyrmion and vortex states, on the other hand, have weights fairly evenly distributed among the four sublattices. This key difference leads to a significant asymmetry in the emergent Kondo impurity couplings for the Kondo hole defect compared to the nearly isotropic vortex and skyrmion two-channel couplings.

\subsection{Dependence on gap size, lattice size and skyrmion radius~\label{supp:scaling}}

\begin{figure}[htb]
\includegraphics[width=0.8\columnwidth]{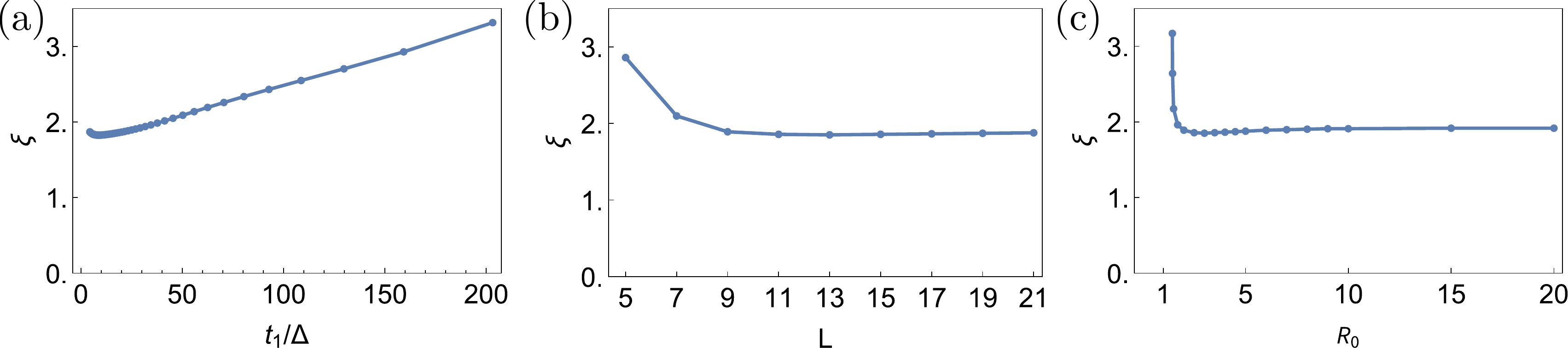}
\caption{Localization length $\xi$ (in number of unit cells) of the type I skyrmion defect state as a function of (a) the inverse of the Kondo insulating gap, controlled by varying $|V|$ from $0.3$ to $1$; (b) the lattice size, $L$, with $R_0=1/4 (L-1)$; (c) the skyrmion radius $R_0$ for a lattice of size $L=21$. For sufficiently large $L$ and $R_0$, the defect state is highly localized with a nearly constant localization length. The parameters used for the plots above are $\mu=-2$, $t_{2}=-2$, $\chi_{1}=0.01$, $\chi_{2}=0.04$ and $|V|=0.5$ for (b) and (c).}\label{fig:locL}
\end{figure}

In this section, we analyze how the localization length of the defect states varies. The localization length, $\xi$ is taken to be $\xi=\sqrt{\text{IPR}}/2$, where the inverse participation ratio, IPR, is given by 
\begin{equation}
\text{IPR}\left[\psi\right]=\frac{1}{\sum_{\mathbf{r},a=A,B,C,D}\left|\psi\left(\mathbf{r},a\right)\right|^{4}}.
\end{equation}
This standard definition of the IPR (in contrast to the modified IPR involving only the $f$ weight defined in the Methods and shown in the previous section) roughly counts the number of sites of any type that the defect state is localized on, with $\xi$ thus being a good proxy for the localization radius. Here, $\xi$ has units of the number of unit cells. The defect states are robust, as long as there is a full insulating gap, and we show in Fig.~\ref{fig:locL}(a) that $\xi$ grows linearly with the gap inverse ($1/\Delta$) for small gaps, but is independent of the gap for larger gaps.  The defect states shown in the previous section are all in this large gap region. 

The localized defect state properties are also mostly independent of system size, as shown in Fig.~\ref{fig:locL}(b) for the type I skyrmion defect. The localization length is mostly independent of system size as long as the system size is sufficiently larger then the localization length itself. In Fig.~\ref{fig:locL}(c), we show that the localization length is also mostly independent of the skyrmion radius as long as $a\ll R_0 < L$, with $a$ the lattice spacing. Typically, the localization length is significantly smaller than the skyrmion radius. The state delocalizes once the skyrmion radius reaches a size comparable to unit cell; this is because the skyrmion is no longer well defined.

\subsection{Defect states away from quadratic band touching~\label{supp:FS}}

Here, we examine the more generic case of the Kondo insulating gap opening up around a full Fermi surface, rather than a single point.  Now there is the possibility of \emph{multiple} defect states in the gap, where in the mean-field zero temperature limit, one of the defect states is half-filled and the others either empty or full. Since our impurity Kondo temperature is larger than the gap, any situation with multiple states in the gap must be treated carefully, as it may be more appropriate to consider the $SU(N)$ Kondo effect for an effectively $N$-fold degenerate multiplet.  The situation would be analogous to the single channel Kondo effect with a crystal field splitting, $\Delta_{CEF}$ between two doublets, where if $T_K \gg \Delta_{CEF}$, there is a $SU(4)$ Kondo effect, but if $T_K \ll \Delta_{CEF}$, there is a $SU(2)$ Kondo effect, with crossover behavior in the intermediate region \cite{coleman2015book}.  In the single channel case, the local moments are always full screened at low temperatures.  In our situation, we can show that there are always still two nearly degenerate channels, meaning that the two extremes for a pair of doublets would be the two-channel $SU(2)$ Kondo model considered in the rest of the paper, or the two-channel $SU(4)$ Kondo model for a small splitting, where the overall filling of the defect state is $3/4$ ($N =4, Q = 3$).  The two-channel $SU(4)$ Kondo model has been less well studied than the $SU(2)$ case, but it is known to have a zero-point entropy of $\frac{1}{2}\ln 3$ for filling $Q =1,3$, and to be described by a $SU(4)_2$ conformal field theory~\cite{parcollet}. The defect state hosts a fractionalized  parafermion that may actually be \emph{more} desirable for quantum computation~\cite{Groenendijk_parafermions}, and this situation should be studied more carefully in future work.

\begin{figure}[htb]
\includegraphics[width=0.95\columnwidth]{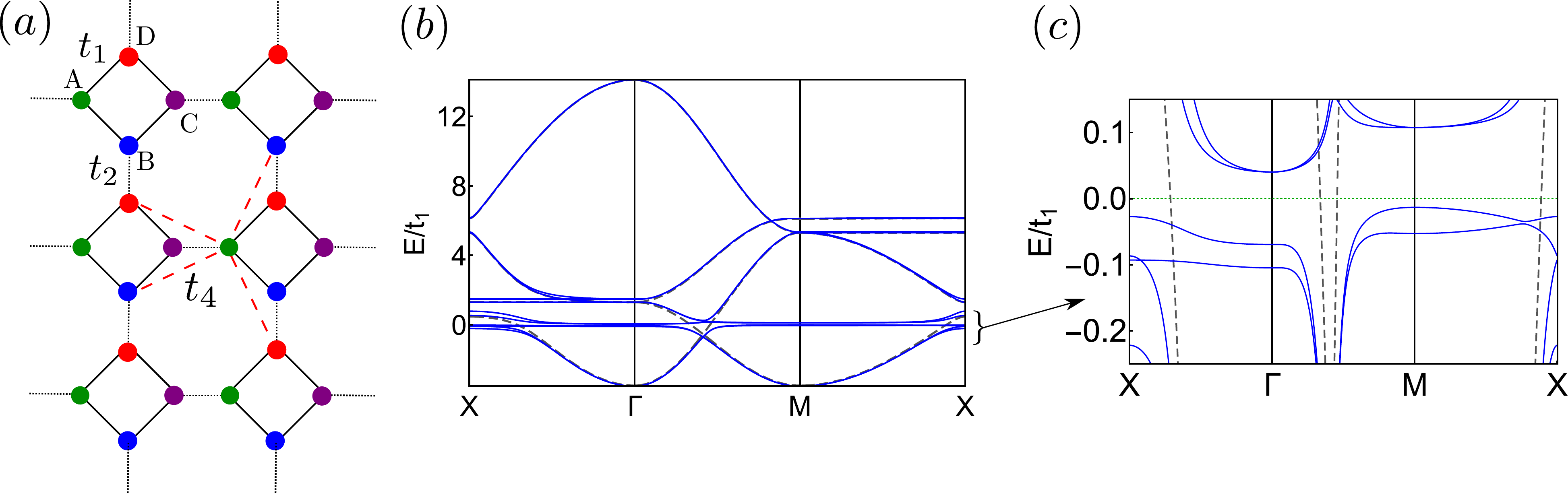}
\caption{(a) Square octagon lattice including the $t_4$ hopping from $A$ to $B$  and $D$ sites of the neighboring unit cells.  Note that all second neighbors within the octagons are connected by $t_4$, although the figure only shows some of these for clarity. This hopping is used to get a full 1/4-filling Fermi surface conduction electron band structure [dashed lines in (b)-(c)]. (b-c) Band structure showing that the AFH order still opens up a Kondo insulator gap at 1/4-filling, with all hybridized bands shown in blue (b) and zoomed in around the gap (c). The parameters used are $\mu=-3.241$, $t_2=-2$, $t_{4}=-1.7$, $\chi_{1}=0.01$, $\chi_{2}=0.04$ and $|V|=0.5$. \label{fig:momentum_space_FS}}
\end{figure}

To obtain a Fermi surface in a square octagon lattice at quarter-filling, we turn on the $t_4$ coupling, as defined in Fig.~\ref{fig:momentum_space_FS}. We chose to turn on the inter-cell hopping $t_4$ instead of a second neighbor intra-cell coupling because $t_4$ gives a broader region of full Kondo insulating gaps as $|V|$ and $|\chi|$ are varied. A fairly large $t_4$ is needed to get a Fermi surface, and the chemical potential $\mu$ is adjusted to enforce quarter filling.

\begin{figure}
\includegraphics[width=0.95\columnwidth]{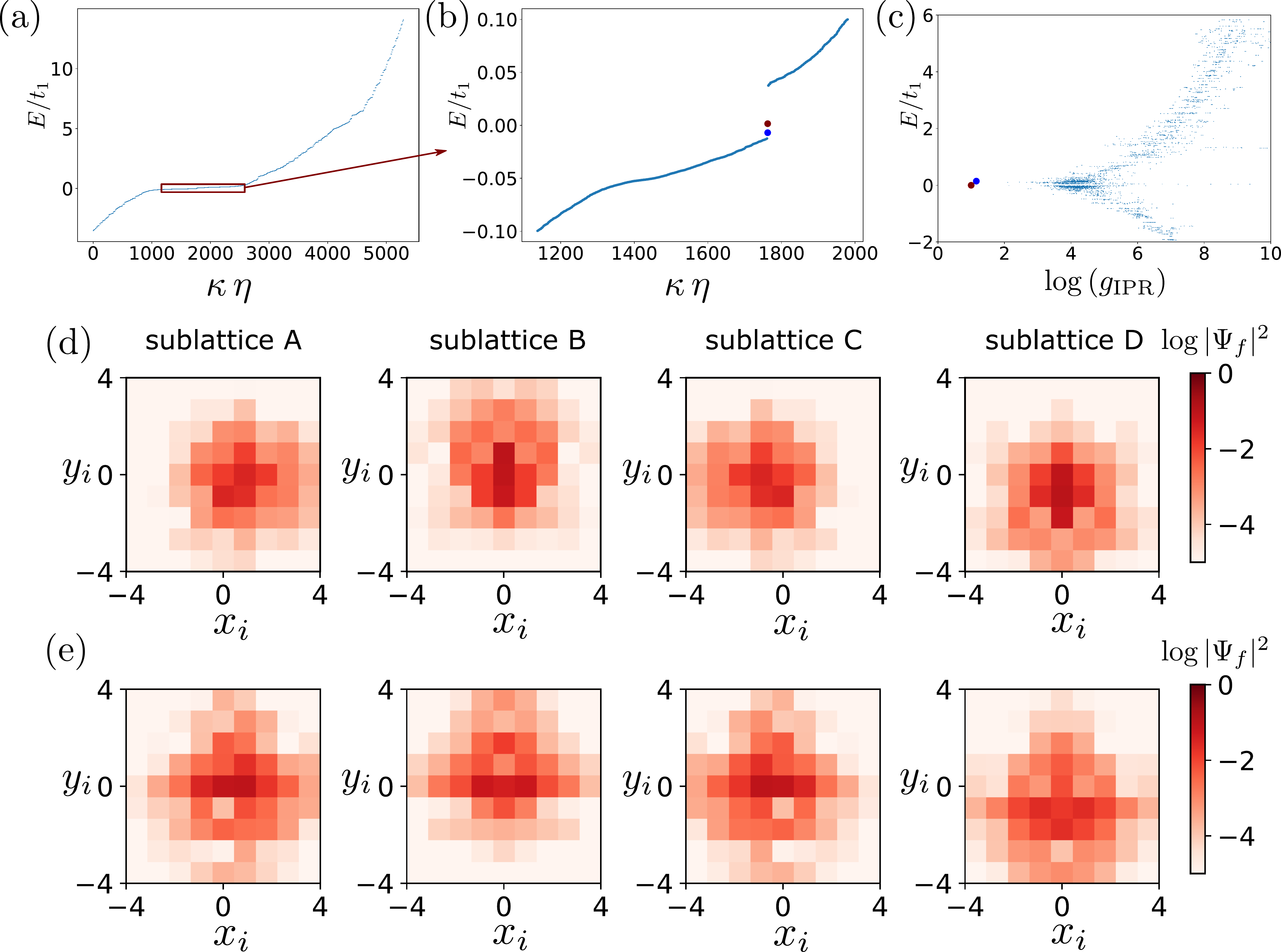}
\caption{Defect states for the type I skyrmion arising from a conduction electron Fermi surface, with parameters $\mu=-3.241$, $t_{2}=-2$, $t_{4}=-1.7$, $\chi_{1}=0.01$, $\chi_{2}=0.04$
and $|V|=0.5$. Each panel is similar to previous figures. (a) shows the energy spectrum, (b) a narrow region near zero energy showing the localized in-gap states with red and blue dots. The $x$-axis is an arbitrary eigenstate label. In (c), the modified IPR for all eigenstates is shown, with the red and blue dots indicating the localized defect states. The upper (red) defect state is half-filled, while the lower (blue) defect state is fully filled. In (d), the $f$-part of the localized half-filled defect state wave function is shown in real space on the four sublattices. In (e), the $f$-part of the lower (fully) filled defect state wave function is shown. The symmetry and localization properties of both states are similar to the other skyrmion defect states considered previously.\label{fig:skyrWF_FS}}
\end{figure}

\begin{figure}
\includegraphics[width=0.95\columnwidth]{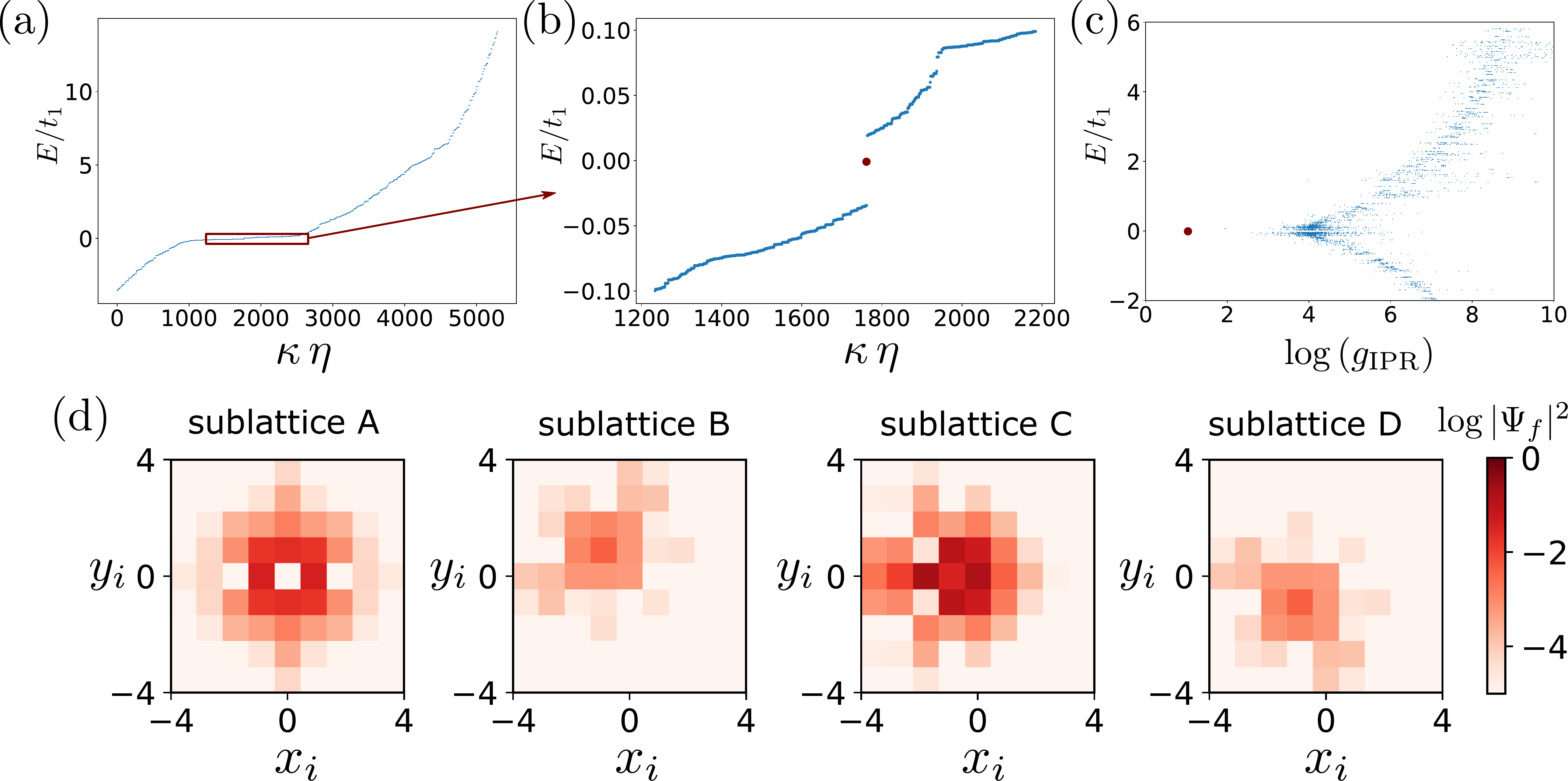}
\caption{Kondo hole defect state arising from a conduction electron Fermi surface, with parameters $\mu=-3.241$, $t_{2}=-2$, $t_{4}=-1.7$, $\chi_{1}=0.01$, $\chi_{2}=0.04$
and $|V|=0.5$. Each panel is similar to previous figures. (a) shows the energy spectrum, (b) a narrow region near zero energy showing the localized in-gap state with a red dot. The $x$-axis is an arbitrary eigenstate label. In (c), the modified IPR for all eigenstates is shown, with the red dot indicating the localized defect state. In (d), the $f$-part of the localized defect state wave function is shown in real space on the four sublattices. Symmetry and localization properties of the defect state are similar to the Kondo hole state arising from the quadratic band touching. No additional defect states appear. \label{fig:defWF_FS}}
\end{figure}

The defects are defined as before, and Figs.~\ref{fig:skyrWF_FS} and \ref{fig:defWF_FS} show the states for the type I skyrmion and Kondo hole, respectively. The Kondo hole case resembles the quadratic band touching case, in that there is still a single defect state in the gap.  By contrast, the skyrmion case has two localized in-gap states for this set of parameters, which are each pseudo-spin degenerate, but are not degenerate with one another. In general, in analogy to the bound states of a delta function potential, we expect that at least one defect state will always appear -- the equivalent of an $s$-wave bound state in momentum space, and there may be multiple defect states.  Depending on the parameters employed, however, some bound states might lie in the band continuum and thus be absent from the gap. This is the case for the Kondo hole, and for some parameter regimes for the skyrmion; the physics is clearly sensitive to microscopic details.  We find the lower state to be $f$ in character, and fully occupied in our mean-field theory, which is consistent with the constraint as the defect state is distributed over all four sublattices. The upper state is also $f$ in character and will be half-occupied in our mean-field theory. We can examine the Kondo interaction for both defect states (see Section \ref{sec:tfres}), considering them, for simplicity, to be half-filled and isolated. We find that they have similar magnitude, $J_I$, and channel asymmetries, $\delta J_I \lesssim 1\%$, ensuring some kind of fractionalization. A more careful treatment is necessary, but should also consider more realistic models and is beyond the scope of this paper.

\section{Gaussian fluctuations - full propagator~\label{GF_prop}}

\begin{figure}
\includegraphics[width=0.99\columnwidth]{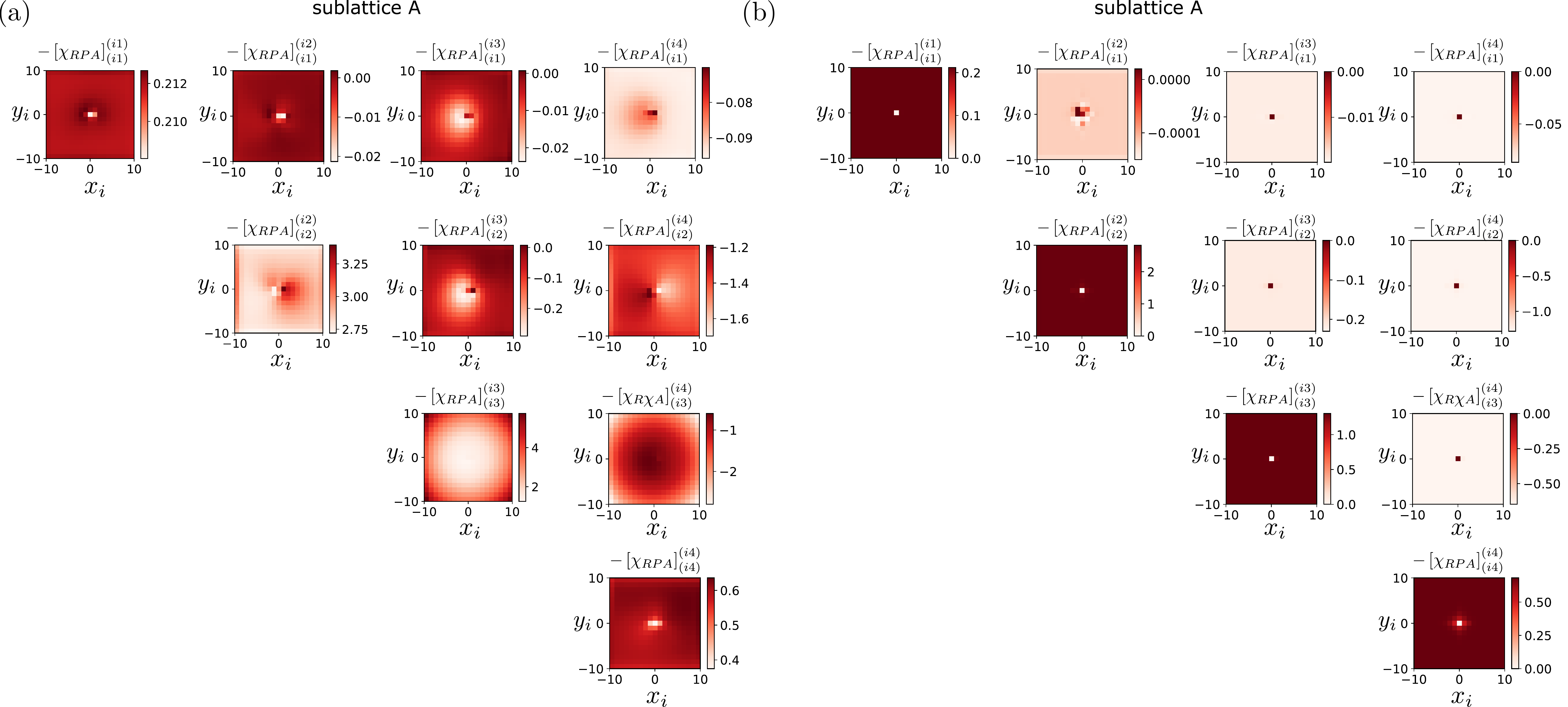}
\caption{RPA fluctuation propagator in real space for the A sublattice for (a) skyrmion and (b) Kondo hole defects, in units of $t_1$. The parameters for the plots were given in Fig.~\ref{fig:skyrWF} and \ref{fig:defectWF}, respectively, with $J_K=0.5 t_1$ for both cases. Only the spatially diagonal parts are shown here, as they dominate, although the full matrix structure is used in our calculations. The angular fluctuations are pronounced and give the dominant contribution to the impurity Kondo coupling.
While the magnitude of the matrix elements is comparable for both cases, the spatial dependence for skyrmion is smooth, with characteristic feature length comparable to skyrmion radius, while the Kondo hole propagator shows abrupt changes in the immediate vicinity of the hole. 
\label{fig:prop_comb}}
\end{figure}

In this section, we show some of our results for the zero frequency fluctuation propagator in real space, which is a key input into the emergent impurity Kondo coupling. This calculation is described in the Methods, and gives $\left[\chi_{RPA}\right]_{(ia\mu)}^{(ja'\nu)}$.  Here, $\mu,\nu = 1,2,3,4$ label the $\delta|V|,|V|\delta \theta, |V|\delta \phi,\delta \lambda$ fluctuation components; $i,j$ labeling the sites; and $a$, $a'$ label the A,B,C,D sublattices.  As the fluctuations interact with conduction and $f$ fermions on-site, the diagonal terms ($i = j$, $a = a'$) are significantly larger than the off-diagonal contributions, and capture most of the physics.  The real-space profiles of these diagonal components of the inverse RPA propagator are shown in Fig.~\ref{fig:prop_comb}(a)-(b) for the skyrmion and Kondo hole, with the parameters used in the main text, and only the $A$ sublattice propagators shown. As the propagator matrix is symmetric, only the upper triangular parts are displayed. The angular fluctuations dominate, with smaller radial and constraint contributions, although the angular fluctuations couple strongly to the constraint fluctuations. This result is not unexpected, as the low energy angular fluctuations are Goldstone modes of the antiferrohastatic order, while the amplitude fluctuations are gapped.  Note that these massless transverse fluctuations do not give divergent interactions with the fermions, as, among other things, the interaction vertices also vanish with $|q|$~\cite{Watanabe2014}. 
Nevertheless, the emergent Kondo coupling $J_I$ mostly comes from the transverse fluctuations and mainly depends on $J_K$ through $|V|$, as opposed to its appearance in $\chi_0$. We can confirm this numerically by artificially giving the transverse fluctuations a large mass and examining the final $J_I$, which is roughly reduced by $\sim 75\%$, indicating the angular fluctuations are responsible for most of the coupling. 
For the more realistic cases of hastatic order in either cubic~\cite{Zhang2018} or tetragonal symmetries~\cite{Chandra2013,Kornjaca2020}, both $\delta \theta$ and $\delta \phi$ (in cubic) or $\delta \phi$ (in tetragonal environments) are only weakly pinned, leading to a small gap that is unlikely to significantly affect any of this physics.

The differences between the skyrmion and Kondo hole cases make it clear that the spatial structure of the RPA propagator stems mostly from the spatially dependent hastatic spinor through the $s_{\mu\xi}^i$ matrix (see Methods). In particular for the skyrmion, the spatial variation has a characteristic length of the order of the skyrmion radius, not the defect localization length. By contrast, for the Kondo hole, the spatial structure is limited to the close vicinity of the hole site, where it changes sharply.

Another aspect of the RPA propagator is the spectrum, where for a stable configuration,  $-\chi_{RPA}$ will have only positive eigenvalues, leading to an antiferromagnetic $J_I$ (see Methods).
Both skyrmion and the Kondo hole cases satisfy this requirement. 
The positivity of the eigenvalues suggests that our parameter choices are not significantly different from the self-consistent real space mean-field values.
For the vortex, however, there is an extensive number of negative eigenvalues.  As vortices are not topologically or energetically stable, this is not surprising.  Nevertheless, these negative eigenvalues have little effect on the final calculation, where $J_I$ is still found to be antiferromagnetic, and the vertices/RPA propagators for different defect types can be ``mixed and matched,'' where the propagator affects mainly the overall magnitude of $J_I$, while the vertices affect the channel asymmetry.

\section{Effective impurity coupling \texorpdfstring{$J_I$}{JI} and channel asymmetry~\label{GF_Kondo}}

In this section, we first examine how the two-channel Kondo impurity coupling scales with the lattice Kondo coupling and lattice size, with results shown for the skyrmion defect discussed in the main text.  We then give a more careful estimate of the impurity Kondo temperature to argue that it should always be greater than the Kondo insulating gap.  Finally, we discuss the temperature scale associated with Fermi liquid formation due to the small channel asymmetry.  All in all, this section details the argument that the Majorana regime should extend from very low temperatures up to nearly the antiferrohastatic ordering temperature.

\subsection{Scaling with system size, \texorpdfstring{$J_K$}{JK}, and \texorpdfstring{$|V|$}{|V|}~\label{sec:scaling}}

\begin{figure}[htb]
\includegraphics[width=0.95\columnwidth]{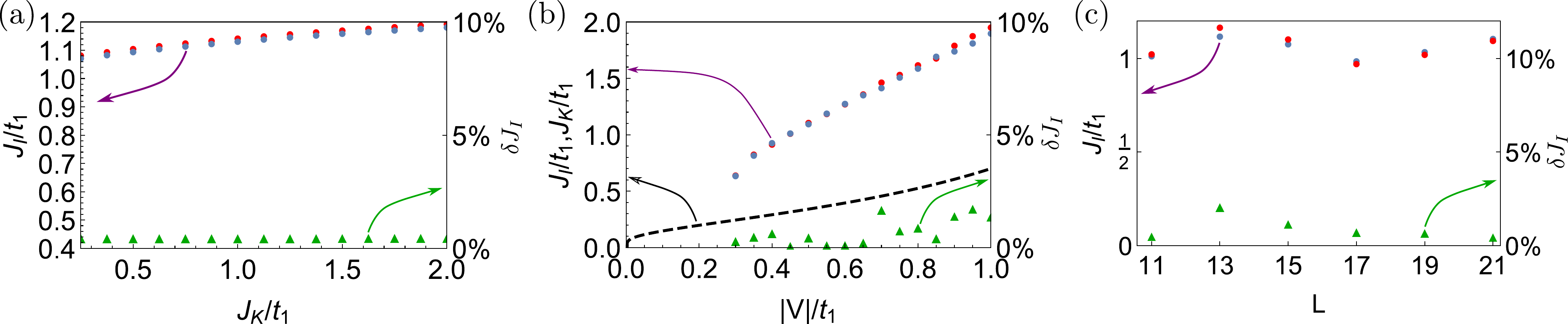}
\caption{Scaling of the emergent impurity Kondo coupling $J_I/t_{1}$ for the skyrmion defect. For all plots, the $J_I$ values for the two channels are shown by the almost overlapping blue and red points, while the relative asymmetry is plotted in green. (a) shows $J_I/t_1$ as a function of $J_K/t_{1}$ while keeping $|V|/t_{1}=0.5$.  $J_I$  depends very weakly on $J_K$, as $J_K$ controls the radial fluctuations that are responsible for a small contribution to the total $J_I$. (b) shows $J_I/t_1$ as a function of $|V|/t_1$, keeping $J_K/t_{1}=0.5$. $|V|$ controls the Kondo insulating gap, which gives the strong dependence here. The dashed black line shows the $J_K/t_{1}$ needed to self-consistently obtain a given $|V|$ for $D=2t_1$. In the region where a gap is opened, $J_I$ is greater than $J_K$. (c) shows $J_I/t_1$ as a function of system size, fixing $J_K=0.5$, $|V|/t_{1}=0.5$ and the skyrmion radius $R_0 = 1/4(L-1)$. The other parameters used in the plots are given in the caption of Fig.~\ref{fig:skyrWF}. $J_I$ is mostly independent of the system size. In all the cases shown, asymmetry remains at or below 1\%. \label{fig:scaling}}
\end{figure}

As we do not self-consistently solve for our mean-field parameters, the hybridization amplitude, $|V|$ and $J_K$ appear as artificially independent parameters.  The dependence of $J_I$ on $J_K$ directly is quite weak, however, and its effect is therefore mainly captured through the dependence on $|V|$.  In Fig.~\ref{fig:scaling} (a), we show the dependence of $J_I$ and $\delta J_I$ on $J_K$ for fixed $|V|$ for the skyrmion defect examined in the main text.  In Fig.~\ref{fig:scaling}(b), we show the dependence on $|V|$ for fixed $J_K$. The impurity Kondo coupling is roughly linear in $|V|$, as long as $|V|$ is large enough to open up a full gap. The asymmetry, $\delta J_I$ has no clear dependence on $|V|$ and is always less than 2\%.  In this sub-figure, we also plot the value of lattice $J_K$ roughly expected to give $|V|$.  We estimate $|V| \sim \sqrt{T_K D}$, where $D$ is half the bandwidth, and $T_{K}=D\exp\left[-D/\left(2J_{K}\right)\right]$.  This estimate assumes a constant density of states for the conduction band \cite{coleman2015book} and leads to $J_K = D/(4 \log D/|V|)$. We take $D = 2 t_1$, for the lower set of bands, although the full $D=4t_1$ gives a somewhat higher $J_K$ estimate that is still significantly lower than the calculated $J_I$.  We expect the experimental $T_K/D \sim .01-.1$ \cite{DordevicMaple_PRL_2001}, which leads to $J_K/t_1 \sim .1-.5$ 
and $|V|/t_1 \sim .1-1$. Note that we have used a fixed $\chi_1,\chi_2$.  These magnitudes ultimately depend on $J_K$ (through the RKKY effect),  so for smaller $J_K$, the $\chi$s will also be smaller and the Kondo insulating gap appears for smaller $|V|$s. The rest of our calculations use $|V| = .5$ and $J_K = .5$,
which is representative of the whole range. Our conclusion is that for any reasonable $J_K/D$, we expect the impurity Kondo coupling $J_I$ to be larger than the lattice coupling $J_K$. 

Finally, in Fig.~\ref{fig:scaling}(c), we show the scaling of $J_I$ with the system size, again for the skyrmion defect with radius $R_0 = 1/4(L-1)$ unit cells, half the lattice size.  $J_I$ is almost constant, but has some small variance, likely due to the changing skyrmion radius slightly affected the details of the defect state. The asymmetry also varies, but always remains $\lesssim 1\%$. For comparison, we studied the system size scaling for the Kondo hole (not shown), and it had significantly less variance, consistent with the fixed size of the Kondo hole.

\subsection{Estimate of the impurity Kondo temperature~\label{sec:TIsizd}}

In the main text, we proposed a simple estimate of $T_I$ based on the heavy nature of the hybridized bands. This estimate found $T_I \approx T_K$, with the estimate being fairly insensitive to the ratio $a=J_I/J_K$. As it is important for $T_I$ to exceed the Kondo insulating gap, $\Delta/2$,
we do a more careful poor man's scaling calculation in this section to show that the impurity will indeed have $T_I > \Delta/D$, as the renormalized coupling $J_I(D)$ diverges in the renormalization group (RG) flow within the heavy portion of hybridized bands before the RG scale reaches the gap. 

\begin{figure}[htb]
\includegraphics[width=0.55\columnwidth]{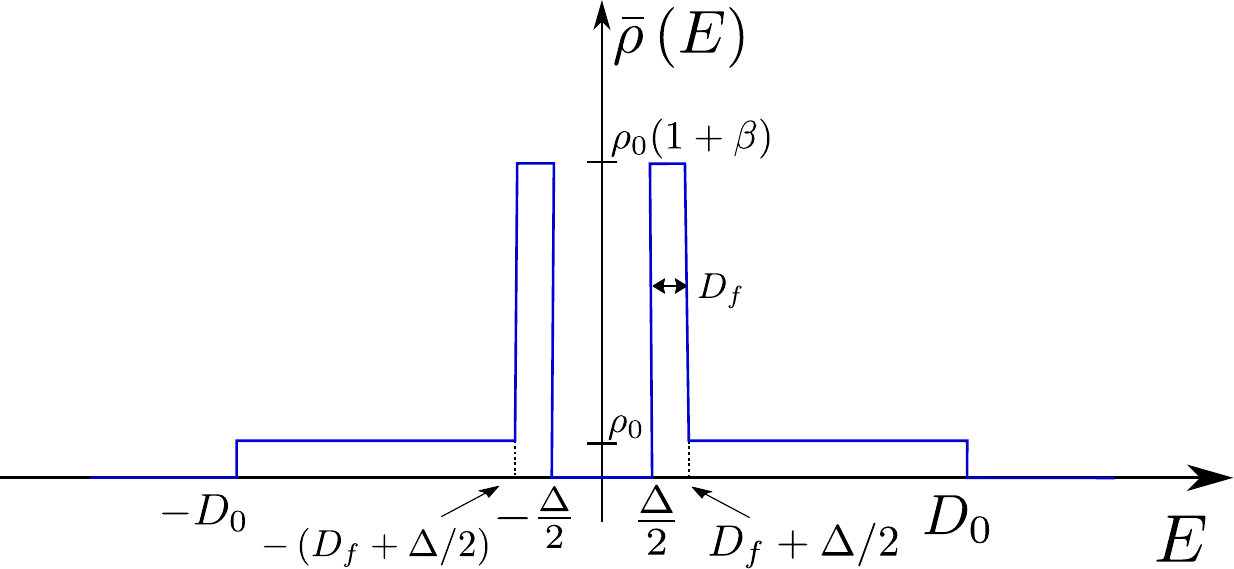}
\caption{Density of states (DOS) $\bar{\rho}(E)$. The  heavy density of states around the gap is modeled by a flat DOS with height $\rho_{0}(1+\beta)$, while for $D_{f}+\Delta/2<\left|E\right|<D_{0}$, the DOS returns to the original $\rho_{0}$. \label{fig:DOS_heavy}}
\end{figure}

We consider an initial conduction density of states $\rho(E)=1/(2D_0)$ between $[-D_0,D_0]$, and then consider a piece-wise constant density of states for the heavy hybridized bands and insulating gap, $\bar{\rho}(E)$, as shown in Fig.~\ref{fig:DOS_heavy}.  This model captures three basic characteristics of hybridized bands, (i) the presence of the Kondo insulating gap $\Delta$, (ii) heavy bands in the vicinity of the gap, of bandwidth $D_f$ and DOS $\beta(1+\rho_0)$, and (iii) only slightly changed light bands away from the hybridization gap. $\beta$ parameterizes the heaviness, and can be estimated from the number of $f$-electrons incorporated into the heavy bands. In our case, four heavy bands are incorporated into four light bands around the Fermi surface, and, to a good approximation, using $\Delta \ll D_0$, $\beta=D_0/D_f\sim D_0/T_K$, as $D_f \sim T_K$ and is thus large (typically 10-100).

We first use the poor man's scaling approximation to find $T_K$~\cite{coleman2015book}, where the flow of the Kondo coupling from high energies ($D_0$) down to the Kondo temperature is,
\begin{align}\label{eq:rgint_lattice}
\int_{D_{0}}^{T_{K}}\frac{\delta J_{K}}{\left[J_{K}(D)\right]^{2}}=-2\rho_{0}\int_{D_{0}}^{T_{K}}\frac{\delta D}{D},
\end{align}
with $T_K$ defined as the energy scale where $J_K(D)$ diverges. Next, we perform the analogous calculation for the impurity, using the density of states of Fig.~\ref{fig:DOS_heavy}. Here we find
\begin{align}\label{eq:rgint_impurity}
\int_{D_{0}}^{T_{I}}\frac{\delta J_{I}}{\left[J_{I}(D)\right]^{2}}=-2\rho_{0}\left[\int_{D_{0}}^{\Delta/2+D_{f}}\frac{\delta D}{D}+\left(1+\beta\right)\int_{\Delta/2+D_{f}}^{T_{I}}\frac{\delta D}{D}\right],
\end{align}
where $T_I$ is the scale at which $J_I(D)$ diverges.

Combining Eqs.~\eqref{eq:rgint_lattice} and Eq.~\eqref{eq:rgint_impurity}, using $J_I=a J_K$, and solving for $T_I$, we find the impurity Kondo temperature:
\begin{equation}\label{eq:TIvsgap}
    \frac{T_I}{\Delta/2}=\left(1+\frac{2D_f}{\Delta}\right)\left(\frac{D_0}{D_f+\Delta/2}\right)^{\frac{1}{1+\beta}}\left(\frac{T_K}{D_0}\right)^{\frac{1}{a(1+\beta)}}.
\end{equation}

If $\beta = D_0/T_K$, the last factor recovers the estimate given in the main text. Here, we can see that for a wide range of realistic parameters, $T_I$ is larger than the gap ($\Delta/2$ is the appropriate comparison).  If we take $D_f \sim \Delta \sim T_K$, $\beta \sim D_0/T_K \gg 1$, we can simplify the weakly changing second and third factors to find
\begin{equation}
 \frac{T_{I}}{\Delta/2}\approx\left(1+\frac{2D_f}{\Delta}\right)\left(\frac{T_K}{D_0}\right)^{\left(\frac{1}{a}-1\right)\frac{T_K}{D_0}}.
\end{equation}
The last factor is of order one, as $T_K/D_0 \sim 0.01-0.1$ and $a \gtrsim 1$,
showing that $T_I \sim 3\Delta/2$ 
for realistic parameters, and that $T_I$ lies within the heavy bands. Notice also that $T_I$ only weakly depends on the ratio of $J_I/J_K$ in the $D_0/T_K \gg 1$ limit. The advantage of the estimate in Eq.~\eqref{eq:TIvsgap} is that $T_I>\Delta/2$ holds regardless of the exact gap size. This is a simple consequence of heavy spectral weight being transferred to just above the gap. In RG terms, for  $a \sim 1$, the light bands present in the unhybridized case contribute similarly to $J_I(D)$ scaling. As we scale down further to the heavy bands, their large DOS leads to the quick divergence of $J_I(D)$. Thus, $T_I$ will tend to lie within the heavy bands as long as $a \sim 1$ and the impurity Kondo effect onsets before the gap scale is reached.  

\subsection{Asymmetry and  the two-channel Kondo crossover scale~\label{sec:asymmetry}}

While in the previous section we discuss the \emph{upper} crossover into Majorana physics in this system, the impurity Kondo scale, $T_I$, in this section we discuss the \emph{lower} crossover scale, $T_x$, governed by the asymmetry, $\delta J_I$.  For perfect channel symmetry, $T_x = 0$, and it grows quadratically in $\delta J_I$, as determined from conformal field theory and numerical renormalization group (NRG) calculations~\cite{Cox1998}, $T_x \sim (\delta J_I)^2 J_I^2/T_I$.  The prefactors can be estimated from NRG calculations, and we take these from [\onlinecite{Pang1991}], which gives,

\begin{figure}[htb]
\includegraphics[width=0.5\columnwidth]{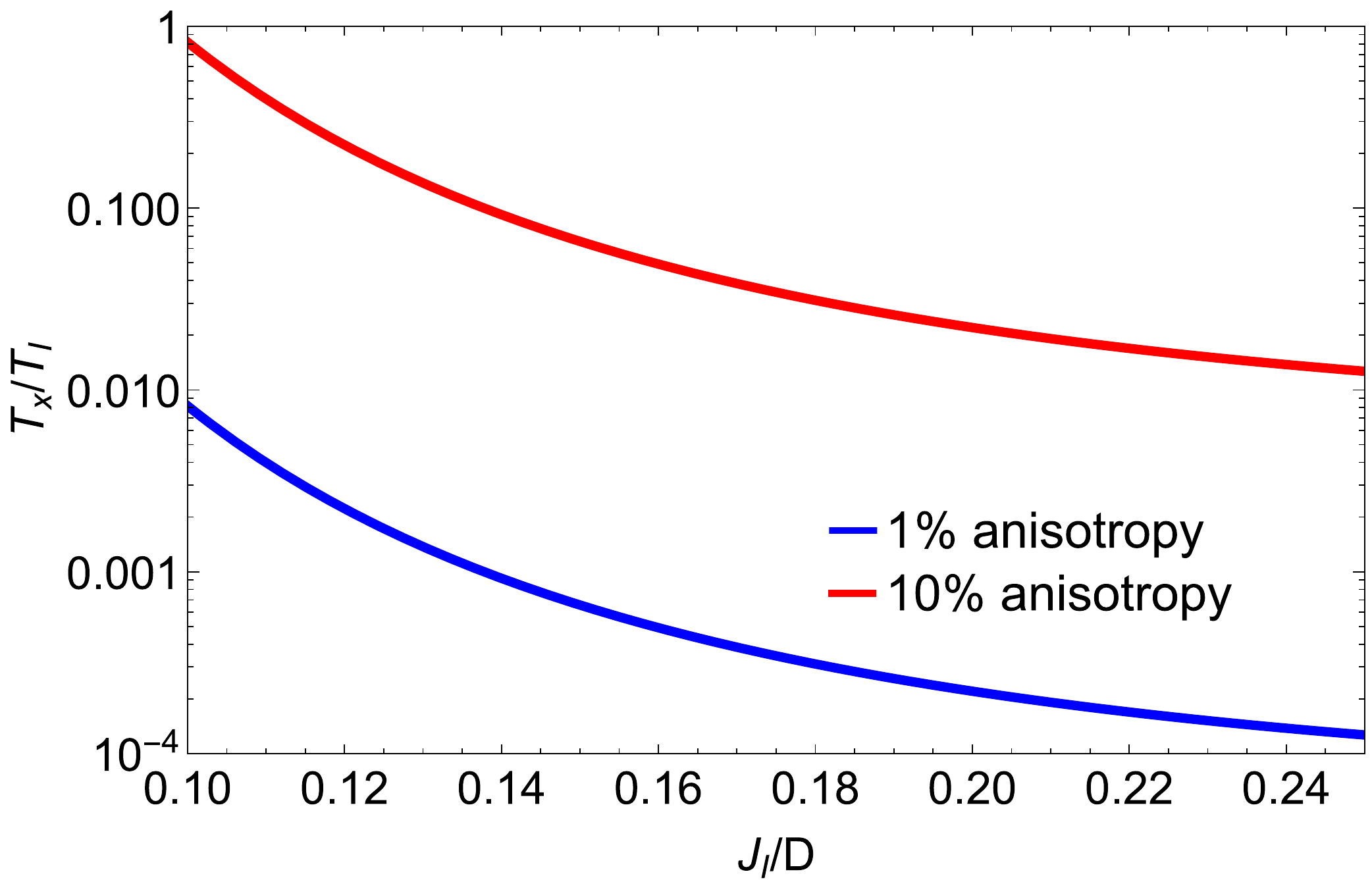}
\caption{Ratio of the two-channel Kondo crossover temperature $T_x$ and the Kondo impurity temperature $T_I$, shown in log scale, as a function of the material-dependent $J_I/D$ ratio (see Eq.~\ref{eq:Tx-estimate}) for 1\% channel asymmetry (blue), 10\% channel asymmetry (red). These are the relevant asymmetries for the skyrmion and Kondo hole defects, respectively. The horizontal axis range corresponds to relevant $J_I/D$ range that comes from $T_K/D \sim 0.01-0.1$. \label{fig:ansize}}
\end{figure}
 
\begin{equation}
   \frac{T_x}{T_I}=\upsilon  \left(2\delta J_I\right)^2 \left(\frac{J_I}{D} \right)^2 e^{\frac{D}{J_I}}, \label{eq:Tx-estimate}
\end{equation}
where the numerical prefactor $\upsilon \approx 0.153/\sqrt{e}$. We plot the ratio $T_x/T_I$ for 1\% and 10\% asymmetry in Fig.~\ref{fig:ansize}, using a log-linear plot.  Here, we consider $J_I \gtrsim .1D$, which captures the range of $T_K/D$ corresponding to real materials.  For $\delta J_I =1\%$, $T_x/T_I$ is well below $0.01$ for the whole range, while for $\delta J_I = 10\%$ the Majorana region is completely eliminated ($T_x = T_I$) for $J_I/D = .01$.  Note that for the skyrmion and vortex defects, $\delta J_I \lesssim 1\%$, meaning the Majorana zero modes should be well defined down to the low temperatures required for relatively error-free braiding. 
The Kondo hole can have asymmetries around 20\%, meaning these defects will be fully screened by the heavy bulk electrons and have no Majorana nature.

\section{Different defect types and robustness to band structure changes~\label{sec:tfres}}

\begin{figure}[htb]
\includegraphics[width=0.95\columnwidth]{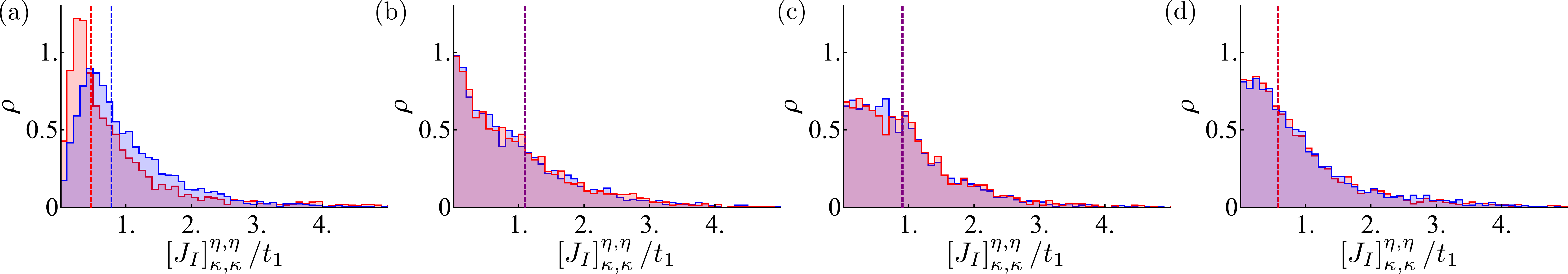}
\caption{Distribution ($\rho$) of the diagonal impurity Kondo couplings $\left[J_{I}\right]{}_{\kappa,\kappa}^{\eta,\eta}/t_1$ for the two Kondo channels, in blue and in red, in the cases of (a) hole and (b) type I skyrmion (c) type II skyrmion and (d) vortex. The vertical lines show the leading $s$-wave coupling, $J_{I}^{(\eta)}$, obtained by averaging the distribution with $J_{I}^{(1)}$ shown in blue and $J_{I}^{(2)}$ in red. In (b)-(d), the two vertical lines are indistinguishable, indicating extremely low asymmetry, while the Kondo hole in (a) has significantly larger asymmetry. The parameters for each case are the same ones used in Fig.~\ref{fig:defectWF}-\ref{fig:vornWF}, respectively, with $J_K=0.5$.  \label{fig:anis_4types}}
\end{figure}

In this section, we give the full results for the impurity Kondo coupling for a wide range of parameter values, demonstrating that the two-channel Kondo physics is robust for different skyrmion and vortex defect types and a range of hopping choices, including the full Fermi surface case. This follows from the results of Section~\ref{real-space-diagon}, which showed that the defect states are robust with respect to parameter changes. In Fig.~\ref{fig:anis_4types}, we show the impurity coupling distributions for four different defect types, similar to Fig.~\ref{fig:Fig4} of the main text. The parameter choices are the same as for the defect states considered in Section~\ref{def_types}. In addition to the channel asymmetric Kondo hole and the nearly channel isotropic type I skyrmion shown in Fig.~\ref{fig:anis_4types} of the main text, the type II skyrmion and vortex show similarly low asymmetries to the skyrmion discussed in the main text.

\begin{table}[htb]
\caption{\label{tab:supp} Emergent Kondo couplings $J_I^{(1,2)}$ in units of $t_1$ and relative channel asymmetry $\delta J_I$ for a range of $f$ hoppings ($\chi_1,\chi_2$), with $t_{2}=-2$ and $|V|=0.5$ and $J_K=0.5$. The hopping $\chi_1$ was varied in the range $0-0.02$, while $\chi_2$ was explored in $0-0.1$ range, with the upper limits such that the system is still gapped. Empty cells indicate the absence of a well-defined localized defect state, as discussed in section \ref{real-space-diagon}. There are two cases for the vortex where we might expect a state, but do not find it to be well-localized; however, we expect that this is an artifact of looking at the vortex in a theory where it is not a stable defect. While the asymmetry for the two skyrmion types and the vortex are comparable and small over the full range of parameters, the Kondo hole shows a significant asymmetry.}
\begin{ruledtabular}
\begin{tabular}{cccccccccccccccccc}
\multicolumn{2}{c}{f-hopping} &  & \multicolumn{3}{c}{Kondo hole}           &  \multicolumn{4}{c}{Type I skyrmion}   & \multicolumn{4}{c}{Type II skyrmion} &  & \multicolumn{3}{c}{vortex}               \\ \hline
$\chi_1$      & $\chi_2$     &  & $J_I^{(1)}$ & $J_I^{(2)}$ & $\delta J_I$ &  & $J_I^{(1)}$  & $J_I^{(2)}$ & $\delta J_I$ &  & $J_I^{(1)}$  & $J_I^{(2)}$  & $\delta J_I$ &  & $J_I^{(1)}$ & $J_I^{(2)}$ & $\delta J_I$ \\ \hline
0 & 0 &  & 0.41 & 0.78 & 31.0\% &  &  &  &  &  &  &  &  &  &  &  &  \\
0 & 0.04 &  & 0.47 & 0.78 & 25.1\% &  & 1.11 & 1.09 & 0.7\% &  &  &  &  &  &  &  &  \\
0 & 0.1 &  & 0.55 & 0.88 & 23.0\% &  & 1.15 & 1.15 & 0.2\% &  &  &  &  &  &  &  &  \\
0.01 & 0.02 &  & 0.45 & 0.76 & 25.3\% &  & 1.07 & 1.07 & 0.1\% &  & 0.90 & 0.92 & 1.1\% &  & 0.61 & 0.60 & 1.1\% \\
0.01 & 0.04 &  & 0.47 & 0.78 & 24.8\% &  & 1.09 & 1.10 & 0.5\% &  & 0.94 & 0.95 & 0.9\% &  & 0.6 & 0.61  & 1.1\% \\
0.01 & 0.06 &  & 0.50 & 0.80 & 22.8\% &  & 1.11 & 1.11 & 0.0\% &  & 0.97 & 0.98 & 0.6\% &  & 0.62 & 0.62 & 0.2\% \\
0.01 & 0.08 &  & 0.54 & 0.83 & 21.4\% &  & 1.13 & 1.13 & 0.1\% &  & 1.00 & 1.01 & 0.5\% &  & 0.58 & 0.59 & 0.8\% \\
0.01 & 0.1 &  & 0.55 & 0.87 & 22.0\% &  & 1.15 & 1.14 & 0.1\% &  & 1.03 & 1.03 & 0.1\% &  & 0.61 & 0.63 & 1.1\% \\
0.02 & 0.02 &  & 0.52 & 0.73 & 16.9\% &  & 1.07 & 1.07 & 0.0\% &  & 0.85 & 0.86 & 0.5\% &  &  &  &  \\
0.02 & 0.04 & & 0.50  & 0.75 & 	19.7\% &  & 1.04 & 1.05 & 0.2\% &  & 0.91 & 0.90 & 0.8\% &  &  &  &  \\
0.02 & 0.06 &  & 0.52 & 0.77 & 19.8\% &  & 1.07 & 1.08 & 0.5\% &  & 0.95 & 0.95 & 0.0\% &  & 0.62 & 0.64 & 0.9\% \\
0.02 & 0.08 &  & 0.54 & 0.80 & 19.2\% &  &  1.11 & 1.09 & 0.9\% & & 0.99 & 0.99 & 0.0\% & & 0.62 & 0.65 &  2.6\%\\
0.02 & 0.1 & & 0.57 & 0.83 & 18.4\% & & 1.10 & 1.09 & 0.4\% & & 1.02 & 1.02 & 0.1\% & & 0.63 & 0.66 & 2.4\%
\end{tabular}
\end{ruledtabular}
\end{table}

We give further results in Table~\ref{tab:supp}, where we consider a range of $f$-hoppings. For all the cases where a localized defect state is present, both types of skyrmions and vortices show similar values of effective impurity couplings ($J_I \approx (0.6-1.1) t_1$) and small channel asymmetry ($\delta J_I \lesssim 1\%$). The Kondo hole in general shows significant asymmetry (mostly $\gtrsim 20\%$), and thus is likely fully screened, while the other defects will host Majorana zero modes for large temperature ranges.

\begin{figure}[htb]
\includegraphics[width=0.78\columnwidth]{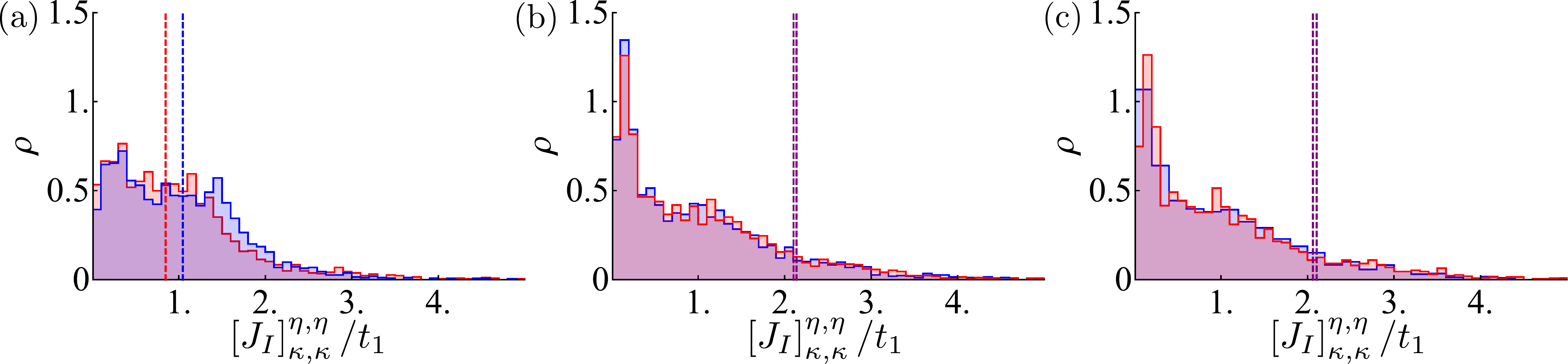}
\caption{Distribution ($\rho$) of the diagonal impurity Kondo couplings $\left[J_{I}\right]{}_{\kappa,\kappa}^{\eta,\eta}/t_1$ for the two Kondo channels (blue and red) in the cases of (a) Kondo hole and (b,c)  skyrmion defect states arising from the full Fermi surface (defect states discussed in Sec.~\ref{supp:FS}), where (b) is the upper, half-filled state and (c) is the lower state, treated as half-filled, although it is actually fully filled. The vertical lines show the leading $s$-wave coupling, $J_{I}^{(\eta)}$, obtained by averaging the distribution. $J_{I}^{(1)}$ is shown in blue and $J_{I}^{(2)}$ in red. The resulting asymmetric behavior is similar to the quadratic band touching case, despite significant changes in the electronic band structure. The parameters for the two cases are the same ones used in Figs.~\ref{fig:skyrWF_FS} and ~\ref{fig:defWF_FS}, respectively, with $J_K=0.5$. The channel asymmetries are (a) 10.9\% for the hole, (b) 0.8\% for the upper defect state and (c) 1.3\% for the lower defect state. \label{fig:anis_FS}}
\end{figure}

Finally, Fig.~\ref{fig:anis_FS} shows the diagonal distribution of impurity couplings for the Kondo hole and the skyrmion in the case of a full Fermi surface (see Section~\ref{supp:FS}). This case is potentially interesting for real materials.  While the coupling size is larger than in the QBT case, this leads to a similar $T_I$ (see main text) with the small channel asymmetry, again putting the skyrmion configuration firmly within the Majorana fan and Kondo hole out of it. 

\end{document}